\documentclass[]{aastex7}

\usepackage{mathrsfs}
\usepackage{amsmath}

\newcommand{\tps}{T_{\rm ps}}
\newcommand{\gcc}{\rm{g\ cm^{-3}}}

\newcommand{\cs}{c_{\rm s}}
\newcommand{\vrad}{v_{r}}
\newcommand{\vphi}{v_{\phi}}

\newcommand{\bz}{B_{z}}

\newcommand{\bph}{B_{\phi}}
\newcommand{\bzeq}{B_{z,\rm{eq}}}
\newcommand{\brs}{B_{r,\mathcal{S}}}
\newcommand{\bphs}{B_{\phi,\mathcal{S}}}

\newcommand{\etaA}{\eta_{\rm AD}}
\newcommand{\etaeff}{\eta_{\rm eff}}

\begin{document}

\title{Development of 1-D non-ideal MHD simulation code towards understanding Long-term Evolution of Protoplanetary Disk \footnote{Footnotes can be added to titles}}

\author[orcid=0000-0002-2163-9677,gname=Yudai,sname=Kobayashi]{Yudai Kobayashi}
\affiliation{Graduate School of Science and Engineering, Kagoshima University, Kagoshima 890-0065,
Japan}
\email[show]{k7983964@kadai.jp}  

\author[orcid=0000-0001-6273-805X,gname=Daisuke, sname=Takaishi]{Daisuke Takaishi} 
\affiliation{Daiichi Institute of Technology}
\affiliation{Graduate School of Science and Engineering, Kagoshima University, Kagoshima 890-0065,
Japan}
\email{d-takaishi@daiichi-koudai.ac.jp}

\author[gname=Yusuke,sname=Tsukamoto]{Yusuke Tsukamoto}
\affiliation{Graduate School of Science and Engineering, Kagoshima University, Kagoshima 890-0065,
Japan}
\email{tsukamoto.yusuke@sci.kagoshima-u.ac.jp}

\author[gname=Shantanu,sname=Basu]{Shantanu Basu}
\affiliation{Canadian Institute for Theoretical Astrophysics, University of Toronto, 60 St. George St., Toronto, ON M5S 3H8, Canada}
\affiliation{Department of Physics and Astronomy, The University of Western Ontario, London, ON N6A
3K7, Canada}
\email{basu@uwo.ca}

\begin{abstract}

We developed a one-dimensional magnetohydrodynamic (MHD) simulation code to investigate the long-term evolution of protoplanetary disks with low computational cost.
In this simulation code, the physical processes necessary for protostellar formation and protoplanetary disk evolution, such as magnetic braking, non-ideal MHD effects, and angular momentum transport due to viscosity, are implemented.
Using this simulation code, we performed the simulations of the long-term evolution of protoplanetary disks starting from the molecular cloud.
Our simulation results suggest that the disk size and mass are a few tens of au and $\sim 0.01 M_\odot$ at $10^5$ years after protostellar formation. These values were relatively consistent with observations.
The disk evolves through magnetic braking, and its radial profiles are consistent with the analytical solutions of previous studies.
Our simulation code will be an important tool for studying the long-term evolution of protoplanetary disks.

\end{abstract}



\section{Introduction}

The recent high-resolution/high-sensitivity observations revealed the nature of the protoplanetary disks around Class 0/I young stellar objects (YSOs) (e.g., eDisk\citep{2023ApJ...951....8O}, VANDAM\citep{2018ApJ...866..161S}, CALYPSO\citep{2019A&A...621A..76M}, CAMPOS\citep{2024ApJ...973..138H}). 
In particular, eDisk \citep{2023ApJ...951....8O} shows that disks around Class 0/I objects do not have ring/gap structures seen in disks around Class II objects.
CAMPOS \citep{2024ApJ...973..138H} shows that the average sizes of disks around Class 0, Class I and flat-spectrum protostars were 59.3 au, 46.0 au and 41.6 au, respectively.
VANDAM \citep{2018ApJ...866..161S} shows that the typical disk mass is less than $0.1 ~M_\odot$ in Class 0/I objects.
These observational studies provide valuable information for understanding the formation and evolution processes of protoplanetary disks over a timescale of a million years.

Theoretically, magnetic fields are thought to play a crucial role in the formation and early evolution of protoplanetary disks, and the formation and early evolution of protoplanetary disks have been investigated mainly using three-dimensional non-ideal magnetohydrodynamic (MHD) simulations.
Non-ideal MHD effects (Ohmic dissipation, Hall effect, and ambipolar diffusion) determine the degree of coupling between the magnetic field and gas. They decrease the efficiency of magnetic braking (the angular momentum transfer due to the magnetic field).
These previous studies show that the non-ideal MHD effects play the central role for the early evolution of the protoplanetary disks, and the disks with sizes of several au to several 10 au form even in molecular clouds with magnetic field strengths comparable to the observations \citep{2011PASJ...63..555M, 2015ApJ...801..117T, 2016A&A...587A..32M, 2017ApJ...838..151T, 2019MNRAS.486.2587W, 2021MNRAS.502.4911X, 2023ASPC..534..317T, 2024ApJ...970...41M}.

These simulation studies have revealed important mechanisms in the formation and evolution of protoplanetary disks, but there is still a major problem. The time scale for disk evolution that can be calculated by simulation is generally around $10^4$ years. On the other hand, the age of observed disks is thought to be around $10^6$ years after the protostellar formation.
The short integration time of the simulation is due to the fact that three-dimensional simulations require a huge amount of computational cost, and it is not expected that it will become possible to calculate the time evolution of the disk up to $10^6$ years in the near future. In other words, there is a huge gap of more than 10 times between the age of the disk in the three-dimensional simulation and the observed age of the disk.

In this study, we have developed a new simulation code that can consider magnetic braking and non-ideal MHD effects, despite being a one-dimensional simulation code. 
%For the formulation and discretization, we extended the non-ideal MHD equations and numerical methods used in \citet{2012A&A...541A..35D} which employ the one-dimensional axisymmetric tihn-disk approximation, by introducing a viscosity diffusion term and sink cells, which enabled us to perform calculations $10^5$ years after protostellar formation.
For the formulation and discretization, we extended the non-ideal MHD equations and numerical methods used in \citet{2012A&A...541A..35D} which employ the one-dimensional axisymmetric thin-disk approximation, by introducing a viscosity diffusion term, which enabled us to perform calculations $10^5$ years after protostellar formation.
Because it is 1D, this code requires dramatically small computational costs. Using this code, we calculate the evolution process of a protoplanetary disk $\sim 10^5$ years after the formation of a protostar. We have confirmed that the disk structure obtained by the simulation reproduces the analytical steady-state solutions of previous studies well.

\section{Method}
\subsection{Basic Equation}
We solve magnetohydrodynamic equations with axisymmetric and thin-disk approximation,
\begin{eqnarray}
    \frac{\partial \Sigma}{\partial t}
    &=& -\frac{1}{r}\frac{\partial}{\partial r}\left( r\Sigma \vrad \right),
    \label{EoC}\\
    \frac{\partial}{\partial t}\left( \Sigma \vrad \right)
    &=& -\frac{1}{r}\frac{\partial}{\partial r}\left( r\Sigma \vrad\vrad\right)
    -\frac{\partial P}{\partial r}+\Sigma g_{\rm r}+\Sigma r\Omega^2
    +\frac{\bzeq}{2\pi}\left(\brs-H\frac{\partial \bzeq}{\partial r}\right),
    \label{EoM}\\
    \frac{\partial L}{\partial t}
    &=& -\frac{1}{r}\frac{\partial}{\partial r}\left(rL\vrad\right)
    +\frac{r\bphs\bzeq}{2\pi}
    +\frac{1}{2\pi r}\frac{\partial}{\partial r}\left(2\pi r^3\alpha \cs H\Sigma\frac{\partial \Omega}{\partial r}\right),
    \label{EoL}\\
    \frac{\partial \bzeq}{\partial t}
    &=& -\frac{1}{r}\frac{\partial}{\partial r}\left(r\bzeq \vrad\right)
    +\frac{1}{r}\frac{\partial}{\partial r}\left(r\etaeff\frac{\partial \bzeq}{\partial r}\right),
    \label{EoB}
\end{eqnarray}
where $\Sigma$ is the surface density of neutral gas, $\vrad$ is the radial velocity, $P$ is the gas pressure, $g_{\rm r}$ is the gravitational acceleration, $H$ is the scale height, $\cs$ is the sound speed , $L\equiv\Sigma\Omega r^2$ is the angular momentum per unit area and $\Omega$ is the angular velocity.
$\bzeq$ is the vertical magnetic field at the midplane.
$\brs$ and $\bphs$ are the radial and azimuthal magnetic fields at the disk surface.

Equation \eqref{EoM} is the equation of motion in the radial direction.
The second, third, fourth, and fifth terms on the right-hand side are the pressure gradient force, gravitational force, centrifugal force, and Lorentz force, respectively.
Note that the Lorentz force term has $H\partial \bzeq/\partial r$ despite the thin-disk approximation (the term $H/r$ is neglected).
This term is needed to form ''magnetic wall'' \citep{2002ApJ...580..987K}.
Equation \eqref{EoL} is the equation for the angular momentum evolution.
The second and third terms on the right-hand side represent the removal of angular momentum by the magnetic field (called magnetic braking) and viscous diffusion.
We describe the model of magnetic braking and viscous diffusion in Section \ref{ss:magnetic braking} and \ref{ss:alpha viscosity}, respectively.
Equation \eqref{EoB} is the induction equation.
The second term on the right-hand side represents magnetic diffusion due to non-ideal MHD effects. We include only ambipolar diffusion by assuming the effective resistivity $\etaeff$ to be $\etaA$.
In this paper, we ignore the Ohmic dissipation and the Hall effect just for simplicity.

The quantities in the basic equations are given as,
\begin{eqnarray}
    H(r)
    &=& \frac{\Sigma(r)}{2\rho(r)},\\
    g_{\rm r}(r)
    &=& 2\pi G \int^\infty_0dr'r'\Sigma_n(r')\mathcal{M}(r,r')
    +\frac{\partial \Psi_{\rm ps}}{\partial r}
    \label{eq:grav}, \\
    \mathcal{M}(r,r')
    &=& \frac{d}{dr}\int^\infty_0dkJ_0(kr)J_0(kr')\nonumber\\
    &=& \frac{2}{\pi}\frac{d}{dr}\frac{1}{r_>}K\left(\frac{r_<}{r_>}\right), \\
    \brs(r)
    &=& -\int^\infty_0dr'r'[\bzeq(r')-B_{\rm ref}]\mathcal{M}(r,r')
    +\frac{\Phi_{\rm ps}}{2\pi r^2}
    \label{eq:brs},\\
    \bphs(r)
    &=& -\frac{2\Phi(r)}{r v_{\rm A,ext}}\Omega(r)\left(
    1+\frac{\Phi(r)}{r^2 v_{\rm A,ext}}\frac{\etaeff(r)}{H(r)\bzeq(r)}
    \right)^{-1}
    \label{eq:bphs},\\
    \Phi(r)
    &=& \int^\infty_0dr'r'\bzeq(r'),
\end{eqnarray}
where $\rho$ is the gas density, $\Phi$ is the magnetic flux, $B_{\rm ref}$ is the uniform external magnetic field of the surrounding medium with density $\rho_{\rm ext}$, and $v_{\rm A, ext}=B_{\rm ref}/\sqrt{4\pi\rho_{\rm ext}}$ is the Alfven velocity in the external medium.
$\mathcal{M}$ is the kernel for the calculation of gravitational and magnetic fields, $J_0$ is the Bessel function of the first kind, $K$ is the Complete Elliptic Integral of the first kind, and the symbols $r_<$ and $r_>$ denote the smaller and larger of $r$ and $r'$, respectively.
$\Psi_{\rm ps}=GM_{\rm ps}/r$ and $\Phi_{\rm ps}$ represent the gravitational potential and magnetic flux of the central star (with mass of $M_{\rm ps}$), respectively.
We calculate the gravitational field $g_{\rm r}$ with the integral approximation for thin disks \citep{1991ApJ...371..296M,1994ApJ...421..561M}.

$\brs$ and $\bphs$ are the radial and azimuthal magnetic fields at the disk surface (indicated by the subscript $\mathcal{S}$).
The radial component $\brs$ is calculated from a potential field assuming force-free and current-free conditions in the external medium, using $\mathcal{M}$.
The gravitational field $g_{\rm r}$ is also calculated using it.
The azimuthal component $\bphs$ determines the magnetic torque.
We implement the models for $\bphs$ according to \citet{1994ApJ...432..720B} and \citet{2012MNRAS.422..261B} (see Section \ref{ss:magnetic braking} and \ref{ss:azimuthal magnetic diffusion} in detail).
$\brs$ and $\bphs$ are masked by $\bzeq$ because we assumed $\brs,\bphs\le\bzeq$.

\subsection{The model for azimuthal magnetic field}
\subsubsection{Ideal MHD limit}\label{ss:magnetic braking}
We adopted the magnetic braking model in \citet{1994ApJ...432..720B} to estimate $\bphs$ in the ideal MHD limit. The torque per unit area exerted on the disk surface is assumed to be,
\begin{eqnarray}
    N_{\rm{mag},\mathcal{S}}
    = -\frac{\Phi}{\pi v_{\rm A,ext}}\Omega\bzeq
    \label{eq:N_mag}.
\end{eqnarray}
The magnetic torque per unit area is $N_{\rm mag}=r\bphs\bzeq/2\pi$. Combined with Equation \eqref{eq:N_mag}, the azimuthal component of the magnetic field is given as,
\begin{eqnarray}
    B_{\phi, \mathcal{S},\rm{MB}} = -\frac{2\Phi}{r v_{\rm A,ext}}\Omega.
\end{eqnarray}
$B_{\phi, \mathcal{S}, \rm{MB}}$ is the azimuthal magnetic field in the ideal MHD limit (i.e., $\etaeff=0$). There is a strong magnetic braking in this limit.

\subsubsection{Azimuthal Magnetic Diffusion}\label{ss:azimuthal magnetic diffusion}
The azimuthal magnetic field of the ideal MHD limit is weakened by magnetic diffusion. We implemented azimuthal magnetic diffusion according to the model in \citet{2012MNRAS.422..261B}. In this model, the azimuthal magnetic field has a drift velocity relative to the gas due to magnetic diffusion in the azimuthal direction.
The drift velocity $v_{B,\phi}$ is given as,
\begin{eqnarray}
    v_{B,\phi}
    % &=& \left[
    % \frac{\etaeff}{B^2}(\nabla\times\mathbf{B})_{\perp}\times\mathbf{B}
    % \right]_{\phi},\nonumber\\
    &\simeq& \frac{1}{2H}\frac{\etaeff}{B^2}\bphs\bzeq,
    \label{eq:vbphi}
\end{eqnarray}
where $B=\sqrt{\brs^2+\bphs^2+\bzeq^2}$.
Then, the angular velocity in Equation \eqref{eq:N_mag} is modified as
\begin{eqnarray}
    \Omega \to \Omega-\frac{v_{B,\phi}}{r}.
\end{eqnarray}
By substituting this to the Equation \eqref{eq:N_mag}, the azimuthal magnetic field including the azimuthal magnetic diffusion $B_{\phi,\mathcal{S},\rm{diff}}$ is given as,
\begin{eqnarray}
    B_{\phi, \mathcal{S}, \rm{diff}}
    &=& B_{\phi,\mathcal{S},\rm{MB}} - \frac{2\Phi}{r^2 v_{\rm A,ext}}v_{B,\phi},\nonumber\\
    &=& -\frac{2\Phi}{r v_{\rm A,ext}}\Omega\left(
    1+\frac{\Phi}{r^2 v_{\rm A,ext}}\frac{\etaeff}{HB^2}\bzeq
    \right)^{-1}.
    \label{eq:bphs_wMD}
\end{eqnarray}
%Equation (\ref{eq:bphs_wMD}) should be numerically calculated because $B=\sqrt{\brs^2+\bphs^2+\bzeq^2}$.
We assume $\bzeq \gg \brs,\bphs$ and $B\sim\bzeq$, thus, Equation \eqref{eq:bphs_wMD} is,
\begin{eqnarray}
    B_{\phi,\mathcal{S},\rm{diff}}
    &= -\frac{2\Phi}{r v_{\rm A,ext}}\Omega\left(
    1
    +\frac{\Phi}{r^2 v_{\rm A,ext}}\frac{\etaeff}{H\bzeq}
    \right)^{-1}.
    \label{eq:bphs_diff}
\end{eqnarray}
%In this paper, the magnetic diffusion on the azimuthal direction is included, and 
$B_{\phi,\mathcal{S},\rm{diff}}$ in the Equation \eqref{eq:bphs_diff} gives the magnetic braking model considered in this paper.

\subsection{Model for viscosity, magnetic resistivity, and Equation of State}
\subsubsection{$\alpha$ viscosity}\label{ss:alpha viscosity}
We adopted the $\alpha$ viscosity model \citep{1973A&A....24..337S} for angular momentum transfer by viscous diffusion.
The parameter $\alpha$ is assumed to be,
\begin{eqnarray}
\alpha=
\begin{cases}
    \frac{1}{Q^4+1} & \text{if }(|\vphi|>2|\vrad|), \\
    0 & \text{otherwise},
\end{cases}
\label{eq:visc}
\end{eqnarray}
where $Q\equiv\kappa\cs/\pi G \Sigma$ is Toomre's $Q$ value \citep{1964ApJ...139.1217T} and $\kappa=\sqrt{4\Omega^2+2r\Omega\partial\Omega/\partial r}$ is an epicyclic frequency.
This models angular momentum transport due to gravitational instability \citep{2010ApJ...713.1143Z,2010ApJ...713.1134Z,2016MNRAS.461.2257K,2018ApJ...865..102T}.

$\vphi=r\Omega$ is the rotational velocity.
By considering Toomre's $Q$ value in the $\alpha$ parameter, the angular momentum transfer due to gravitational instability is modeled as a viscous evolution.
Because gravitational instability should develop only in the disk, $\alpha$ is set to 0 in regions where $|\vphi|<2|\vrad|$ is satisfied, i.e., when the radial accretion is dominant.

In this paper, we consider two models: one is with $\alpha$-viscosity ($\alpha$-viscosity is considered only in the disk; named "alphaGI") and the other is without it ($\alpha=0$ in all regions; named "non-alphaGI").

\begin{table}[ht]
    \caption{Simulation models}
    \label{tb:model}
    \centering
    \begin{tabular}{lr} \hline
        model & $\alpha$ viscosity in disk \\ 
        \hline
        alphaGI & $\alpha=\frac{1}{Q^4+1}$ \\
        non-alphaGI & $\alpha=0$ \\
        \hline
    \end{tabular}
\end{table}

\subsubsection{Magnetic resistivity}
The magnetic resistivity of ambipolar diffusion $\etaA$ is calculated with the analytic model of \citet{1983ApJ...273..202S},
\begin{eqnarray}
    \etaA &=& \frac{B^2}{4\pi\gamma \rho C \sqrt{\rho}},
    \label{eq:etaA}
\end{eqnarray}
where $\gamma=3.5\times10^{13}\ \rm{cm^3\ g^{-1}\ s^{-1}}$ is the drag coefficient arising from momentum exchange in ion-neutral collisions and $C=3\times10^{-16}\ \rm{g^{1/2}\ cm^{-3/2}}$ is a constant that describes the balance between the cosmic-ray ionization and the gas-phase recombination.
In this resistivity model, only gas-phase recombination is considered, and the adsorption of charged dust particles is ignored. We will perform calculations using a more realistic model (e.g \cite{2022ApJ...934...88T}) that is improved in this regard in the future study.

\subsubsection{Equation of State}
We adopted the barotropic equation of state (EOS) \citep{2000ApJ...531..350M}. The gas pressure at the midplane $P_{\rm mid}$ is given as,
\begin{eqnarray}
    P_{\rm mid}
    = \rho c^2_{s, {\rm iso}}
    \left(1+\left(\frac{\rho}{\rho_1}\right)^{2/3}\right)
    \left(1+\left(\frac{\rho}{\rho_2}\right)^{2/3}\right)^{-1}\left(1+\left(\frac{\rho}{\rho_2}\right)^{2/5}\right),
    \label{EoS}
\end{eqnarray}
where $c_{s,{\rm iso}}=1.9\times10^4\ \rm{cm\  s^{-1}}$ is the isothermal sound speed, $\rho_1=1.1\times10^{-13}\ \gcc$ and $\rho_2=3.8\times10^{-11}\ \gcc$.

By assuming vertical hydrostatic equilibrium, the pressure at the equatorial plane $P_{\rm eq}$ is given as,
\begin{eqnarray}
    P_{\rm eq} 
    = \frac{\pi}{2}G\Sigma^2 
    + P_{\rm ext} 
    + \frac{1}{8\pi}\left(\brs^2+\bphs^2\right)
    + \frac{GM_{\rm ps}\Sigma}{4r^2}\frac{H}{r},
    \label{EoS2}
\end{eqnarray}
%where $P_{\rm ext}=9.4\times10^{-12}\ {\rm g\ cm^{-1}\ s^{-2}}$ is the external pressure.
where $P_{\rm ext}=0.1(\pi G \Sigma_{c,0}^2/2)$ is the external pressure and $\Sigma_{c,0}$ is the central surface density of the initial molecular cloud.
The fourth term on the right-hand side represents the tidal squeezing due to the central object with the mass $M_{\rm ps}$.
In our simulation code, the midplane density $\rho$ is calculated using the Newton-Raphson method from Equation \eqref{EoS} and \eqref{EoS2} by assuming $P_{\rm eq}=P_{\rm mid}$.

\subsection{Initial conditions}
We assume that the initial cloud core is in hydrostatic equilibrium along the vertical direction.
The initial profiles of surface density $\Sigma(r)$ and angular velocity $\Omega(r)$ are given as,
\begin{eqnarray}
    \Sigma(r)
    &=& \frac{\Sigma_{c,0}}{\sqrt{1+(r/R_1)^2}}\frac{1}{\sqrt{1+(r/R_2)^6}},\\
    \Omega(r)
    &=& \frac{2\Omega_{c,0}}{\sqrt{1+(r/R_1)^2}+1}\frac{1}{\sqrt{1+(r/R_2)^6}},
    % \Sigma(r)
    % &=& \frac{\Sigma_{c,0}}{\sqrt{1+(r/R)^2}},\\
    % \Omega(r)
    % &=& \frac{2\Omega_{c,0}}{\sqrt{1+(r/R)^2}+1},
\end{eqnarray}
where $R_1=1.1\times10^4\ \rm{au}$ approximately equals the Jeans length of the initial cloud core ($L_{\rm Jeans}=1.2\times10^4\ \rm{au}$).
To isolate the molecular cloud and fix the core mass to $\sim 1.0\ M_\odot$, we impose the outer cutoff with the steep radial power law of $r^{-3}$ outside of $R_2\simeq1.6\times10^4\ \rm{au}$.

$\Sigma_{c,0}=3.0\times10^{-2}\ \rm{g\ cm^{-2}}$ is the central surface density and $\Omega_{c,0}=0.40\ \rm{km\ s^{-1}\ pc^{-1}}$ is the central angular velocity.
The initial core mass within $R_1$ is $1.0\ M_\odot$, the core size is $1.1\times10^4\ \rm{au}$, the ratio of thermal energy to gravitational energy is $\sim 0.41$, and the ratio of rotational energy to gravitational energy is $\sim 0.036$.

The initial profile of vertical magnetic field $\bz(r)$ is assumed to be,
\begin{eqnarray}
    % \bz(r) = \frac{2\pi\sqrt{G}}{\mu_0}\Sigma(r),
    \bz(r) &=& \bz^{'}(r) \left(1+\left( \frac{B_{\rm ref}}{\bz^{'}(r)} \right)^6 \right)^{1/6}, \\
    \bz^{'}(r) &=& \frac{2\pi\sqrt{G}}{\mu_0}\Sigma(r),
\end{eqnarray}
where $\mu_0=4.0$ is the initial mass-to-flux ratio of the molecular cloud core. 
The terms enclosed in brackets are attached so that the strength of the magnetic field approaches the external magnetic field, $B_{\rm ref}$, with a sufficiently large radius.We assume that the critical mass-to-flux ratio is $(\Sigma/\bz)_{\rm critical}\equiv1/(2\pi\sqrt{G})$ \citep{1978PASJ...30..671N}. $\bz^{'}(0)=31\ \rm{\mu G}$.
Within the cloud core ($r<R_2$), the mass-to-flux ratio $\mu=(\Sigma/\bz)_{\rm core}/(\Sigma/\bz)_{\rm critical}$ is constant for the radius.
Outside the molecular core ($r>R_2$), the mass-to-flux ratio is subcritical because the lower limit of $\bzeq$.

We choose the external density $\rho_{\rm ext}\simeq1.2\times10^{-21}\ \gcc$ and external magnetic field $B_{\rm ref}\simeq 16\ \mu\rm{G}$.

\subsection{Grid structure and Sink cell}
We adopted logarithmic spacing grid cells of 
\begin{eqnarray}
    % {\zeta}_{i}
    % &=& \zeta_1 \epsilon^{-1} \exp{(i\log(1+\epsilon))},
    % \label{eq:init_grid}
    \Tilde{r}_i = \Tilde{r}_1 \epsilon^{-1} \exp{(i\log(1+\epsilon))},
\end{eqnarray}
where $\Tilde{r}_i$ is the position of the $i$-th cell interface and $\Tilde{r}_1$ is the position of the 1st cell interface.
The subscript $i$ is from 0 to $N$, where $N$ is the number of cells ($N=100$ in the initial grid).
$\epsilon=0.045$ is a parameter that determines the cell size.
The position of the 1st cell interface $\Tilde{r}_1$ is calculated using $\epsilon$ as,
\begin{eqnarray}
    \Tilde{r}_1 = \frac{\epsilon}{(1+\epsilon)^N-1}\Tilde{r}_N,
\end{eqnarray}
where $\Tilde{r}_N=8.8\times10^4\ \rm{au}$ $(=8R_1)$ is the position of the outermost cell interface.
The position of the $i$-th cell center $r_i$ is calculated using $\Tilde{r}_i$ and $\Tilde{r}_{i+1}$ as,
\begin{eqnarray}
    r_i
    &=& \frac{2}{3}\frac{{\Tilde{r}}_i^2 + \Tilde{r}_i \Tilde{r}_{i+1}+{\Tilde{r}}^2_{i+1}}{\Tilde{r}_i+\Tilde{r}_{i+1}}.
    \label{eq:grid}
\end{eqnarray}
Our simulation code adopted a reflection boundary at the origin and a constant-volume boundary at the outermost cell.

The grid cells are remapped so that the Jeans length is always resolved into at least 10 cells during the time evolution. 
When $\Delta r$ of the innermost cell becomes $1\ \rm{au}$ and the gas density exceeds $\rho\simeq3.8\times10^{-11}\ \gcc$ (this value is the density at which the second collapse occurs), the sink cell is introduced.
After introducing the sink cell, the inner boundary is the outgoing boundary and the grid is remapped to a high-resolution grid with $N=200$ cells.

In the high-resolution grid, $\Tilde{r}_1$ and $\Tilde{r}_N$ are fixed.
$\epsilon$ is redefined as,
\begin{eqnarray}
    \epsilon = \mathcal{R}^{-\frac{1}{N+1}}-1,
\end{eqnarray}
where $\mathcal{R}=\Tilde{r}_1/\Tilde{r}_N$.The $i$-th cell interface of the high-resolution grid is given as,
\begin{eqnarray}
    \Tilde{r}_i &=& \Tilde{r}
    _1(1+\epsilon)^{i-1},
    \label{eq:new_zeta}\\
    \Tilde{r}_0 &=& 0.
\end{eqnarray}
The position of the $i$-th cell center $r_i$ is given by Equation \eqref{eq:grid}, using $\Tilde{r}_i$ in Equation \eqref{eq:new_zeta}. This grid structure has a resolution of $\Delta r\sim0.06 r$ at the 1st cell.

The sink cell is treated as a point source of mass $M_{\rm ps}$ and magnetic flux $\Phi_{\rm ps}$. $M_{\rm ps}$ and $\Phi_{\rm ps}$ are the sum of the mass and magnetic flux outgoing from the 1st cell interface at each time step, respectively. Therefore, the mass is conserved in this simulation.

\section{Results} \label{sec:results}
In this paper, the results of alphaGI are adopted as the fiducial model.

\subsection{Prestellar collapse phase} \label{sub:prestellar collapse phase}
\begin{figure}[ht]
    \centering
    \includegraphics[width=0.5\linewidth]{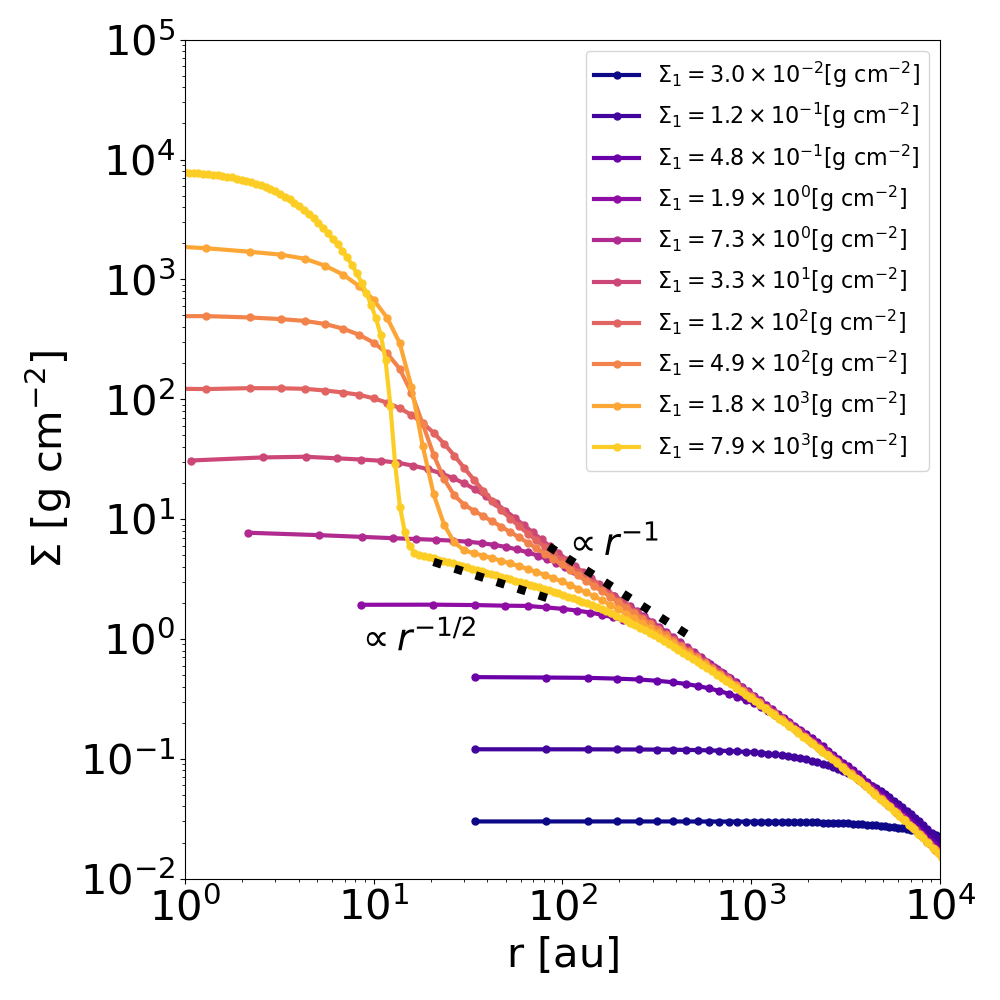}
    \caption{The radial profile of surface density. Color shows the time evolution from the navy plot ($\Sigma_1=3.0\times10^{-2}\ \rm{g\ cm^{-2}}$) to yellow plot ($\Sigma_1=7.9\times10^3\ \rm{g\ cm^{-2}}$). The two black dotted lines represent power laws of $r^{-1}$ (isothermal contraction, \citet{1969MNRAS.145..271L}) and $r^{-1/2}$ (rarefaction wave, \citet{1977ApJ...214..488S}).}
    \label{fig:sigma_cloud}
\end{figure}

Figure \ref{fig:sigma_cloud} shows the time evolution of the surface density.
In the pre-collapse phase, the surface density at the center ($\Sigma_1$) increases with time, so we use the surface density at the center as a time reference.

When $\Sigma_1 \lesssim 7.3\ \rm{g\ cm^{-2}}$, the system is in an isothermal contraction phase, characterized by a central region where $\Sigma(r) \propto r^0$ and an outer region where $\Sigma(r) \propto r^{-1}$.
Once $\Sigma_1 \gtrsim 7.3\ \rm{g\ cm^{-2}}$, the gas in the central region enters adiabatic evolution. This temporarily halts gravitational collapse due to the gas pressure gradient force, allowing a pressure-supported first core to form \citep{1969MNRAS.145..271L, 2000ApJ...531..350M}.
In the final stage of the first core ($\Sigma_1 \sim 7.9 \times 10^{3}\ \rm{g\ cm^{-2}}$), rarefaction waves propagate in the outer region ($r > 10\ \rm{au}$) of the first core, causing the profile to flatten ($\Sigma(r) \propto r^{-1/2}$). This corresponds to Shu's late solution \citep{1977ApJ...214..488S}.

% 図に r^-1 とr^-1/2のべきが欲しい。

\begin{figure}[ht]
    \centering    \includegraphics[width=0.5\linewidth]{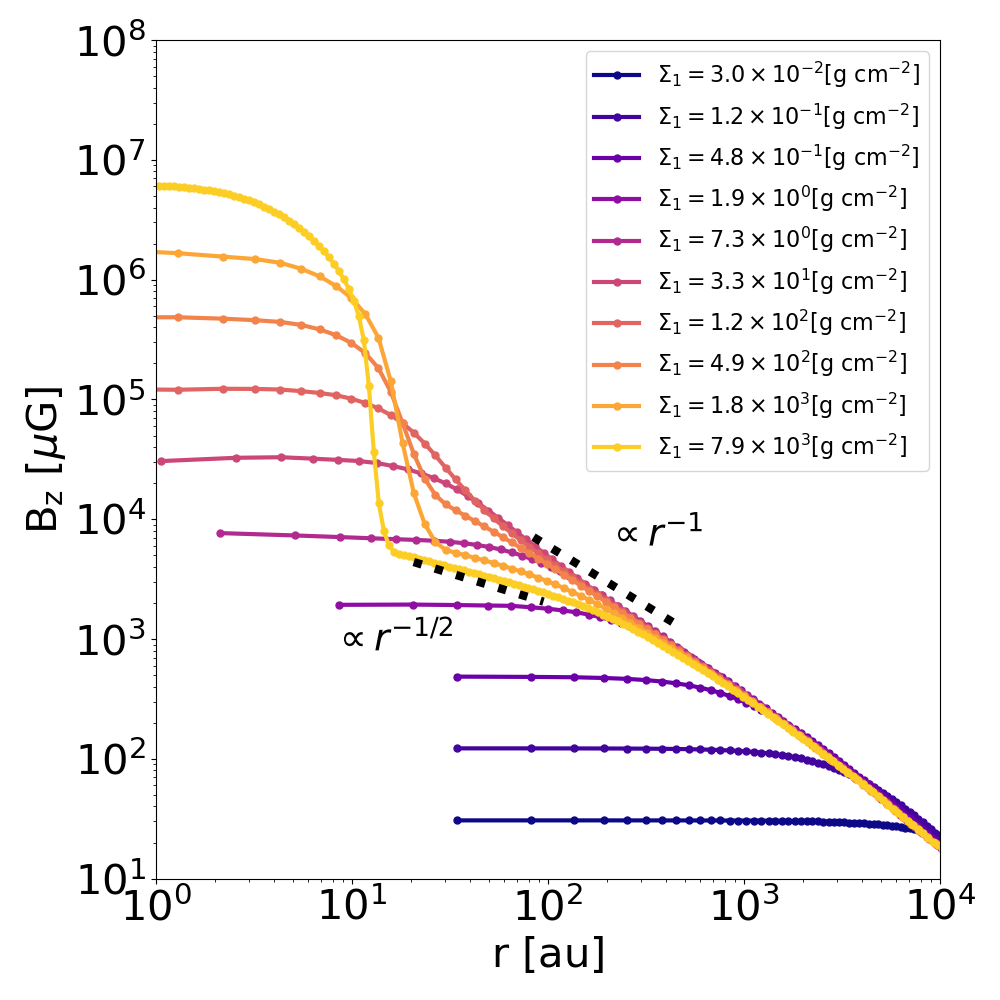}
    \caption{The radial profile of the vertical magnetic field. Color is the same as figure \ref{fig:sigma_cloud}. The two black dotted lines are the same power law as the surface density.}
    \label{fig:bz_cloud}
\end{figure}

Figure \ref{fig:bz_cloud} shows the time evolution of the vertical magnetic field $\bzeq$.

In the prestellar collapse phase, the profile of the vertical magnetic field is characterized by the same features as the surface density profile.
For example, at $\Sigma_1 \lesssim 7.3\ \rm{g\ cm^{-2}}$, the system is in an isothermal contraction phase, and $\bz(r)$ is proportional to $r^0$ in the central region and proportional to $r^{-1}$ in the outer region.
Also, in the final stage of the first core ($\Sigma_1 \sim 7.9 \times 10^3\ \rm{g\ cm^{-2}}$), rarefaction waves propagate in the outer region of the first core ($r > 10\ \rm{au}$), and the profile flattens ($\bz \propto r^{-1/2}$).
These results indicate that the system is evolving under ideal MHD conditions (i.e., $\mu \propto  \Sigma/\bz=\rm{const}$).

% Figure \ref{fig:vn_cloud} shows the radial profile of the radial velocity.
% The first core formation is more evident from this figure.
% The radial velocity is close to zero and a shock forms at $r\sim10\ \rm{au}$.
% %Immediately after protostar formation (the yellow plot), the profile has a bump due to the magnetic wall outside of the first core ($r\sim 6\ \rm{au}$).
% \begin{figure}[ht]
%     \centering
%     \includegraphics[width=0.5\linewidth]{cloud_scale_Rin2.0e+00au/infall_velocity.png}
%     \caption{The radial profile of radial velocity. Color is the same as figure \ref{fig:sigma_cloud}.}
%     \label{fig:vn_cloud}
% \end{figure}

\begin{figure}[ht]
    \centering
    \includegraphics[width=0.5\linewidth]{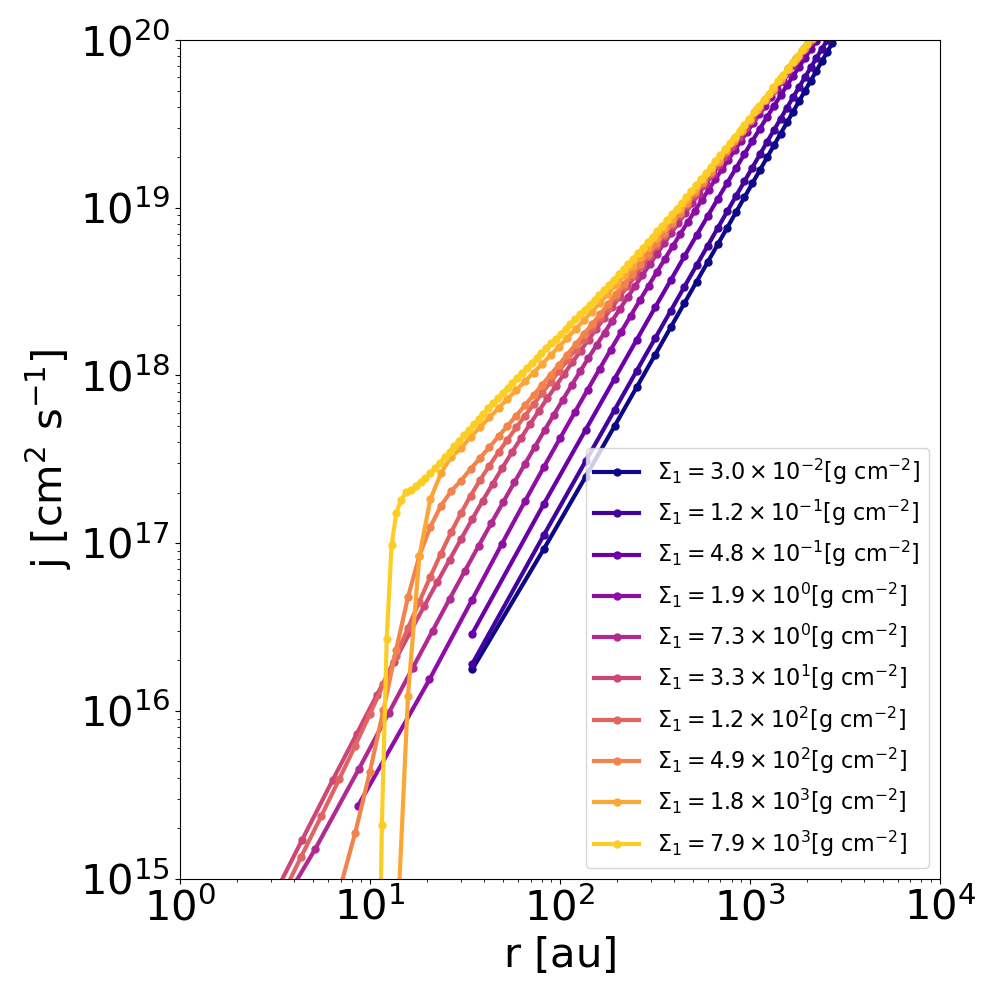}
    \caption{The radial profile of specific angular momentum. Color is the same as figure \ref{fig:sigma_cloud}.}
    \label{fig:am_cloud}
\end{figure}

Figure \ref{fig:am_cloud} shows the time evolution of the specific angular momentum $j\equiv r^2\Omega$.
At $\Sigma_1 \lesssim 7.3\ \rm{g\ cm^{-2}}$, the specific angular momentum retains its initial profile. This indicates that the angular momentum is almost conserved. In other words, the influence of magnetic braking is limited at this stage.

On the other hand, at the final stage of the first core ($\Sigma_1 \sim 7.9 \times 10^3\ \rm{g\ cm^{-2}}$), the specific angular momentum significantly decreases in the central region. This indicates that once the first core is formed, magnetic braking efficiently removes angular momentum within it, leading to a nearly rotation-less state.
This is consistent with earlier multidimensional simulations of angular momentum evolution \citep{2000ApJ...528L..41T,2015ApJ...801..117T,2015ApJ...810L..26T}.

%However, immediately after protostar formation (the yellow plot), the bump forms around 10 au.
%This bump is due to the diffusion of angular momentum due to gravitational instability.
%Outside of the first core ($10\rm{au}<r<100\rm{au}$), $\Omega\propto r^{-1}$ because the angular momentum is conserved with the specific angular momentum $j\equiv\Omega r^2$ and $\Sigma\propto r^{-1}$ in the prestellar collapse phase (figure \ref{fig:sigma_cloud}).

\subsection{Disk evolution in protostellar phase} \label{sub:disk evolution}

\subsubsection{Long-term evolution of the disk and its radial structure}
In this section, we focus on the formation and evolution of the protoplanetary disk after protostellar formation.
$\tps$ represents the time elapsed since protostellar formation.
We define the disk as the region where the rotation velocity and sound velocity are larger than the radial velocity
($|\vphi|>2|\vrad|$ and $\cs>2|\vrad|$). Similar definition is used in the previous studies \citep{2011MNRAS.413.2767M,2013A&A...554A..17J,2016A&A...587A..32M}.
The sound velocity criterion ($\cs>2|\vrad|$) was added to avoid underestimating the disk size due to pressure bumps.

 %We define the disk as a rotationally dominant and $Q>1$.
%The definition of disks is,
%\begin{eqnarray}
    %\frac{|\vphi|}{\mathrm{min}(|\vrad|,\cs)} &>& 2,\\
    %|\vphi|&>&2.0|\vrad|. \label{disk_def1} \\
    %|\vrad|&<&0.1\cs, \label{disk_def2} \\
    %Q&=&\frac{\cs\Omega_K}{\pi G \Sigma} > 1, \label{disk_def3}
%\end{eqnarray}
%where $\vphi=r\Omega$ is the rotational velocity and $\Omega_K=\sqrt{GM_{\rm ps}/r^3}$ is the Keplerian angular velocity.
%$Q$ is Toomre's $Q$ value using the Keplerian rotation.

\begin{figure}[ht]
    \centering
    \includegraphics[width=0.5\linewidth]{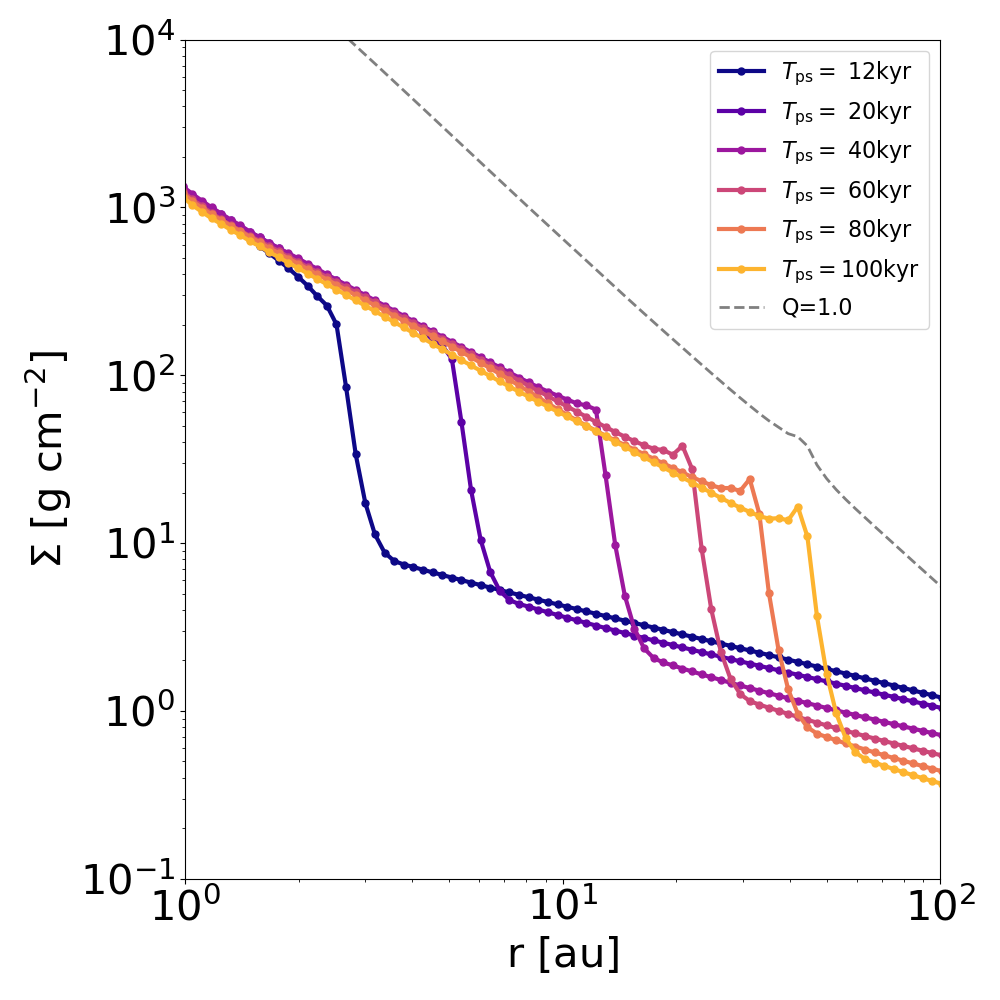}
    \caption{The radial profile of surface density. The color shows the time evolution from the navy plot to the orange plot. $\tps$ is the age of protostar.The navy plot line shows the profile as protostar forms ($\tps=1.2\times10^4$ years) and the orange plot shows $\tps=10^5$ years later. The dashed line shows the $Q=1$ disk at $\tps=10^5$ years.}
    \label{fig:sigma_disk}
\end{figure}

Figure \ref{fig:sigma_disk} shows the time evolution of the surface density after the protostellar formation.
At $\tps=1.2\times10^4$ years, there is a sudden increase in surface density at $r\sim 3$ au, which is a shock. Inside this region is a rotationally supported protoplanetary disk. In our simulation setup, the protoplanetary disk formed at a size of several au about $10^4$ years after protostellar formation (see Figure \ref{fig:t_rdisk}).
The radius of the protoplanetary disk increases with time. Meanwhile, the radial profile looks power law, and its power remains unchanged after disk formation. At $\tps=10^5$ years, the disk size is $R_{\rm d}\sim40$ au. At $r=R_{\rm d}$, a bump is formed by the shock caused by gas accretion from the envelope.

The gray dashed line indicates the surface density of the disk with $Q=1$ at $\tps =10^5$ years. Inside the disk ($r<R_{\rm d}$), $\Sigma<\Sigma_Q$, indicating that the disk is gravitationally stable.

\begin{figure}
    \centering
    \includegraphics[width=0.5\linewidth]{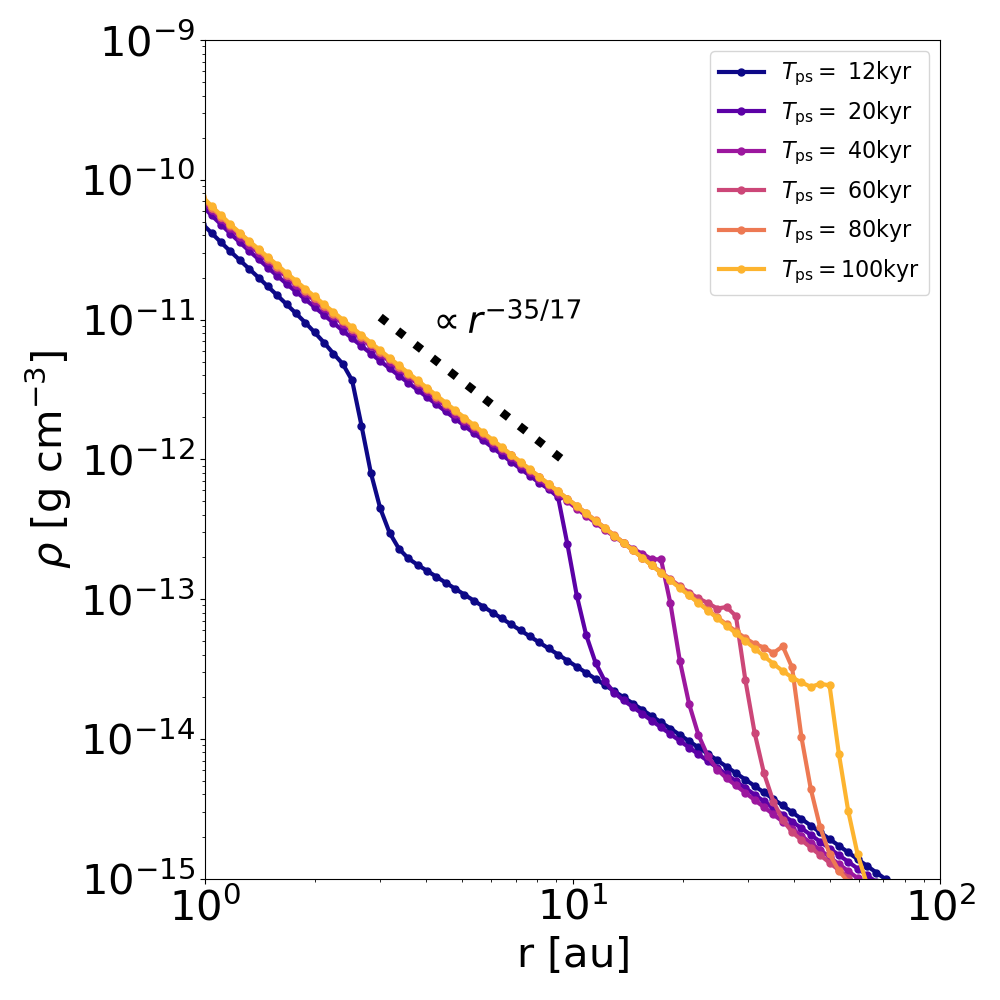}
    \caption{The radial profile of density.Color is the same as figure \ref{fig:sigma_disk}.The black dotted line shows the power law of the solution from \citet{2024PASJ...76..674T}.}
    \label{fig:rho_disk}
\end{figure}

Figure \ref{fig:rho_disk} shows the time evolution of the density. As with the surface density profile, the density profile also follows a characteristic power law. In fact, this power law can be well reproduced by the steady-state disk model proposed by \citet{2023PASJ...75..835T} and \citet{2024PASJ...76..674T}. The equations for obtaining the power law are given as,
\begin{eqnarray}
    && -2\pi r \vrad \Sigma = \dot{M}_{\rm d},\label{eq20:Mdot}\\
    && \Omega_{K} = \sqrt{\frac{GM_{\rm ps}}{r^3}},\label{eq20:EoM}\\
    && H_{\rm d}=\frac{\cs}{\Omega_{K}},\label{eq20:H}\\
    && \Sigma r\Omega_{K} \vrad = -r\frac{\bz\bph}{\pi},\label{eq20:angmom}\\
    && \bz\vrad = -\frac{\etaA}{r}\bz,\label{eq20:bz}\\
    && \etaA\frac{\bph}{H_{\rm d}} = \bz\vphi\label{eq20:bphi},
\end{eqnarray}
where $H_{\rm d}$ is the scale height of the disk, $M_{\rm ps}$ is the protostar mass, and $\dot{M}_{\rm d}$ is the mass acrretion rate in the disk, which is assumed to be constant.
$\Sigma=\sqrt{2}H_{\rm d}\rho$.
$\etaA$ is the magnetic resistivity of ambipolar diffusion \citep{1983ApJ...273..202S}, given by Equation \eqref{eq:etaA}.
We calculated the sound speed $\cs$ using barotropic EOS ($P\propto\rho^{7/5}$). 

Here, Equation \eqref{eq20:bphi} has been modified from \citet{2024PASJ...76..674T} to be consistent with the magnetic braking model discussed in this paper (Equation \eqref{eq:vbphi}). 
The Equation \eqref{eq:vbphi} describes that $\bphs$ (toroidal magnetic field at the disk surface) is determined by the balance between the generation of the magnetic field due to Kepler rotation and the diffusion of the magnetic field in the rotation direction. This model has been used in e.g.,  \citet{2002ApJ...580..987K} and \citet{2012MNRAS.422.2737W} for the studies of one-dimensional gravitational collapse.

On the other hand, \citet{2021AAS...23732507X} and \citet{2023PASJ...75..835T} suggested that $\bphs$ is determined by the balance between the generation of the magnetic field due to vertical shear motion and the diffusion of the magnetic field, and \citet{2024PASJ...76..674T} modeled this result.
In our study, however, we used the model of \citet{2002ApJ...580..987K}. Therefore we have employed the Equation \eqref{eq20:bphi} to be consistent with the Equation \eqref{eq:vbphi}.

By solving the equations, we obtain the power law of $\rho\propto r^{-35/17}$. The black dotted line shows this power law and agrees well with the simulation results.

\begin{figure}[ht]
    \centering
        \includegraphics[width=0.5\linewidth]{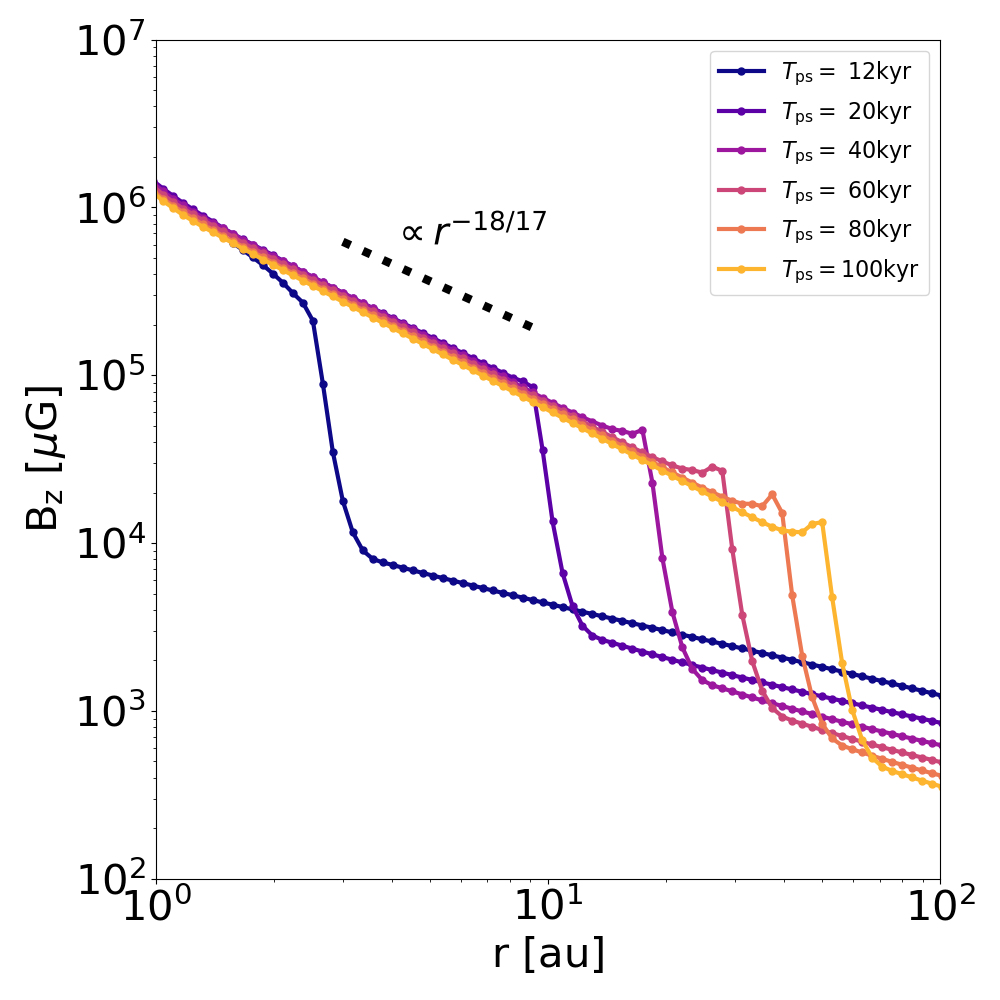}
    \caption{The radial profile of the vertical magnetic field.Color is the same as figure \ref{fig:sigma_disk}.The black dotted line shows the power law of the solution from \citet{2024PASJ...76..674T}. 
    }
    \label{fig:bz_disk}
\end{figure}

Figure \ref{fig:bz_disk} shows the time evolution of the vertical magnetic field after the protostellar formation.
As with the density profile, the vertical magnetic field also follows a characteristic power law.
This power law can also be well reproduced by the steady-state disk model.
The power law calculated from the disk model (Equation \eqref{eq20:Mdot} to \eqref{eq20:bphi}) is $\bz \propto r^{-18/17}$. The black dotted line shows this power law and agrees very well with the simulation results.

% Figure \ref{fig:vn_disk} shows the radial profile of the radial velocity.
% $\tps>1.2\times10^4$ years, in the disk, $\vrad\sim0$ and gas are supported by the centrifugal force.
% Shock forms outside of the disk, and the disk is growing with time.
% %The bump due to the gas accretion forms at the disk edge ($\tps>10^4$ years).
% \begin{figure}[ht]
%     \centering
%     \includegraphics[width=0.5\linewidth]{disk_evolution_until_1.0e+05yr_Rin2.0e+00au_scale100au_wAnalytic/infall_velocity.png}
%     \caption{The radial profile of radial velocity. Color is the same as figure \ref{fig:sigma_disk}.}
%     \label{fig:vn_disk}
% \end{figure}

\begin{figure}[ht]
    \centering
    \includegraphics[width=0.5\linewidth]{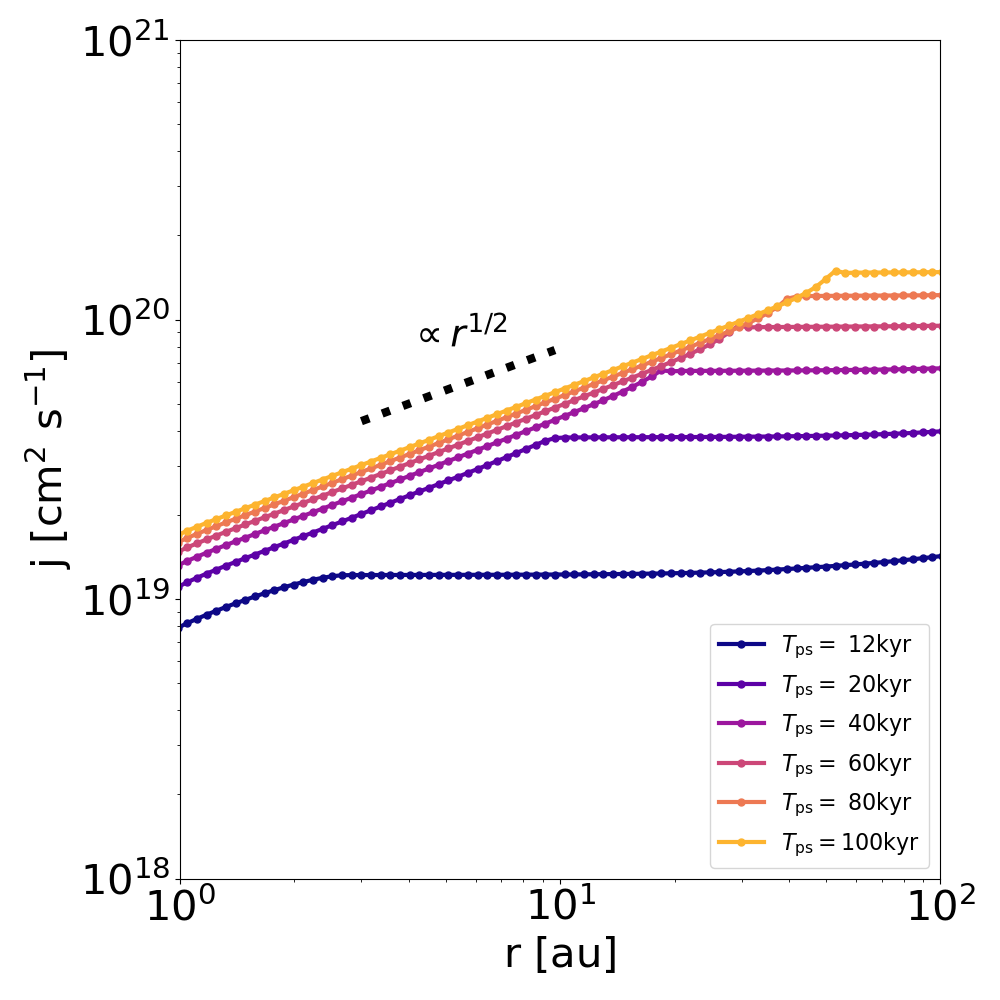}
    \caption{The radial profile of the specific angular momentum. Color is the same as figure \ref{fig:sigma_disk}. The black dotted line shows the power law of Keplerian rotation.}
    \label{fig:am_disk}
\end{figure}

Figure \ref{fig:am_disk} shows the time evolution of the specific angular momentum.
The disk region is characterized by the specific angular momentum profile of Keplerian rotation ($j\propto r^{1/2}$).
On the other hand, the envelope is characterized by the specific angular momentum conservation ($j\propto r^{0}$).

The important feature here is that there is no jump in the angular momentum profile at the boundary between the disk and the envelope. This is a characteristic structure when the disk evolves due to magnetic braking. If the disk evolves due to viscous accretion, the angular momentum at the outer edge of the disk generally does not match that of the accretion flow (see \citet{2024PASJ...76..674T} for more detail).

\begin{figure}[ht]
    \centering
    \includegraphics[width=0.5\linewidth]{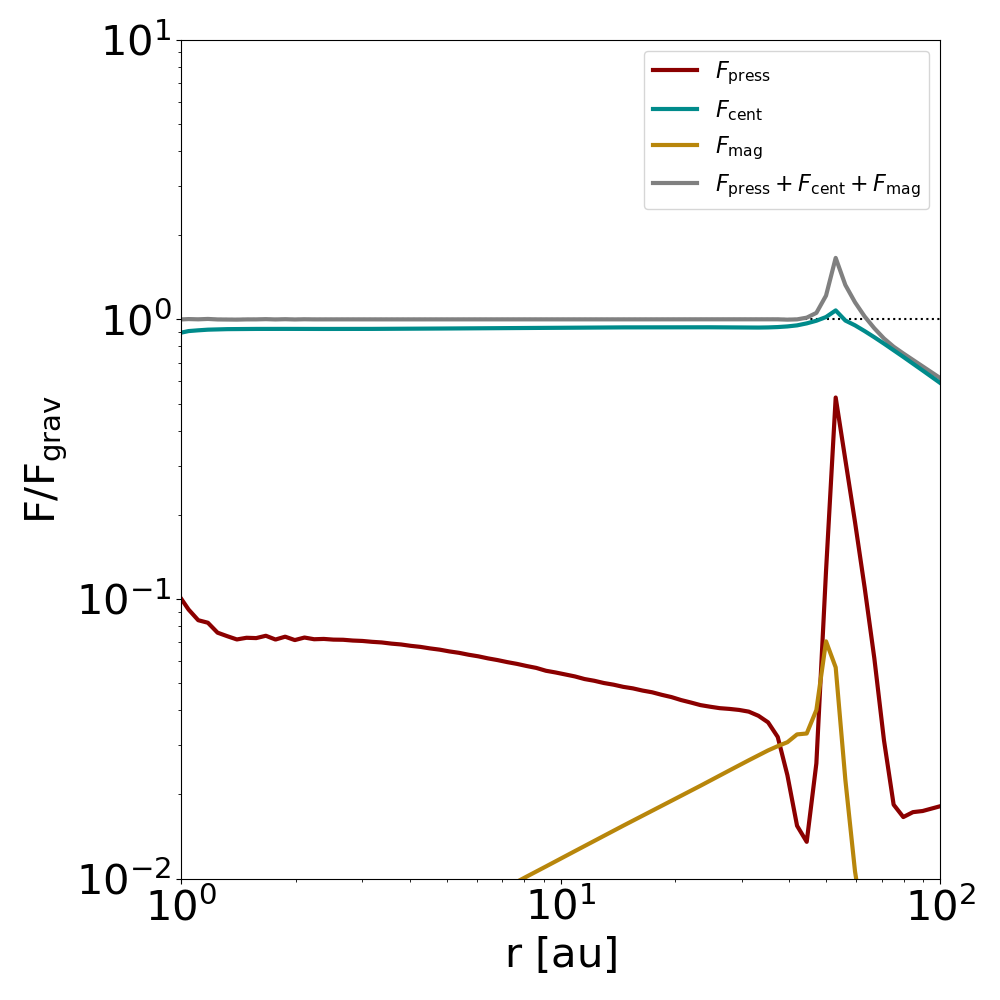}
    \caption{The radial profile of pressure gradient force (red), centrifugal force (cyan), and Lorentz force (gold) normalized by the gravity at $\tps=10^5$ years. The gray line shows the total force.}
    \label{fig:force}
\end{figure}

Figure \ref{fig:force} shows the radial profile of the pressure gradient force (red line), the centrifugal force (cyan line), and the Lorentz force (gold line) normalized by gravity at $10^5$ years after protostellar formation.
Within the disk ($r<R_{\rm d}$), the gas is mainly supported by centrifugal force ($F_{\rm cent}\sim 0.9F_{\rm grav}$), and both the pressure gradient force and the Lorentz force play a minor role in the force balance.

\subsubsection{Time evolution of mass and size of the disk}
% Figure \ref{fig:t_mass} shows the mass profile of the protostar $M_{\rm ps}$ (red line), disk $M_{\rm d}$ (cyan line), and envelope $M_{\rm env}$ (gold line) as the function of time.
% The envelope mass is defined as the initial core mass minus the mass of the protostar and disk ($M_{\rm env}=M_{\rm core, initial}-(M_{\rm ps}+M_{\rm d})$).

% After protostar formation, the protostar mass rapidly increases.
% After disk formation ($\tps>2.0\times10^3$ years), the protostar growth becomes slower than $\tps<2.0\times10^3$ years.
% At $\tps=7.0\times10^4$ years, the protostar mass is larger than the envelope mass.
% Immediately after disk formation, the disk mass increases rapidly.
% In the early phase of disk evolution ($\tps\sim2.0\times10^3$ years), the disk mass is large relative to the protostar and Toomre's $Q$ value is close to 1.
% Thus, the disk mass increases with oscillations.
% After $\tps=7.0\times10^4$ years, the disk mass growth slows and is nearly constant with time.
% The envelope mass tends to decrease with time.
% At $\tps=10^5$ years, $M_{\rm ps}\sim0.57M_\odot$, $M_{\rm d}\sim0.12M_\odot$ and $M_{\rm env}\sim0.35M_\odot$.
% $\sim55\%$ of the initial core mass falls to the protostar and the envelope mass decreases to $\sim33\%$.

\begin{figure}[ht]
    \centering
    \includegraphics[width=0.5\linewidth]{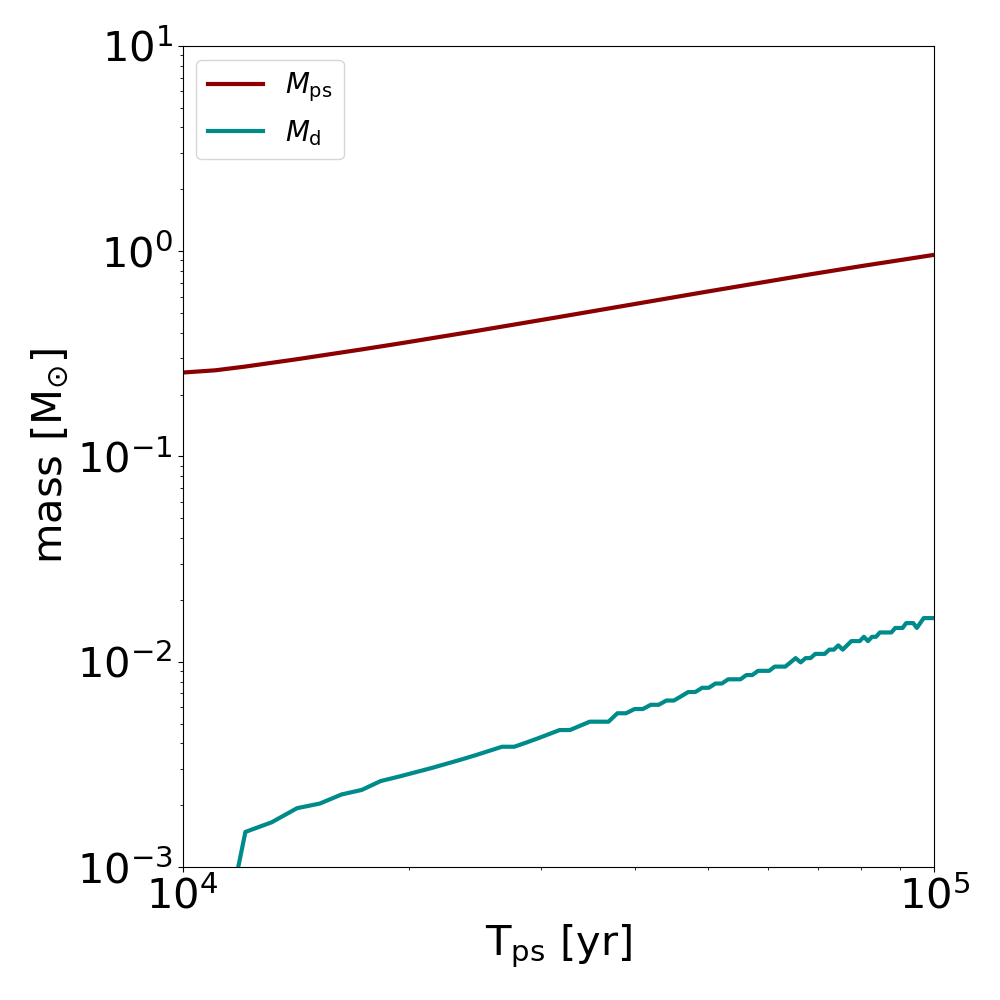}
    \caption{The protostar mass and disk mass as a function of time. Red and cyan lines show the protostar and disk mass, respectively. The vertical axes show the mass and the horizontal axes show time from protostellar formation $\tps$.}
    \label{fig:t_mass}
\end{figure}

Figure \ref{fig:t_mass} shows the time evolution of the mass of the protostar $M_{\rm ps}$ (red line) and the disk $M_{\rm d}$ (cyan line).
At the time of disk formation, the mass of the central star was approximately $M_{\rm ps} \sim 0.2 M_\odot$, and it grows to $M_{\rm ps} \sim 1 M_\odot$ over $10^5$ years.
On the other hand, the mass of the disk was approximately $M_{\rm d} \sim 10^{-3} M_\odot$ at the formation epoch and $M_{\rm d} \sim 1.6\times10^{-2} M_\odot$ at the end of the simulation. Therefore, the disk mass is less than a few percent of the protostar mass throughout the simulation period.

\begin{figure}[ht]
    \centering
    \includegraphics[width=0.5\linewidth]{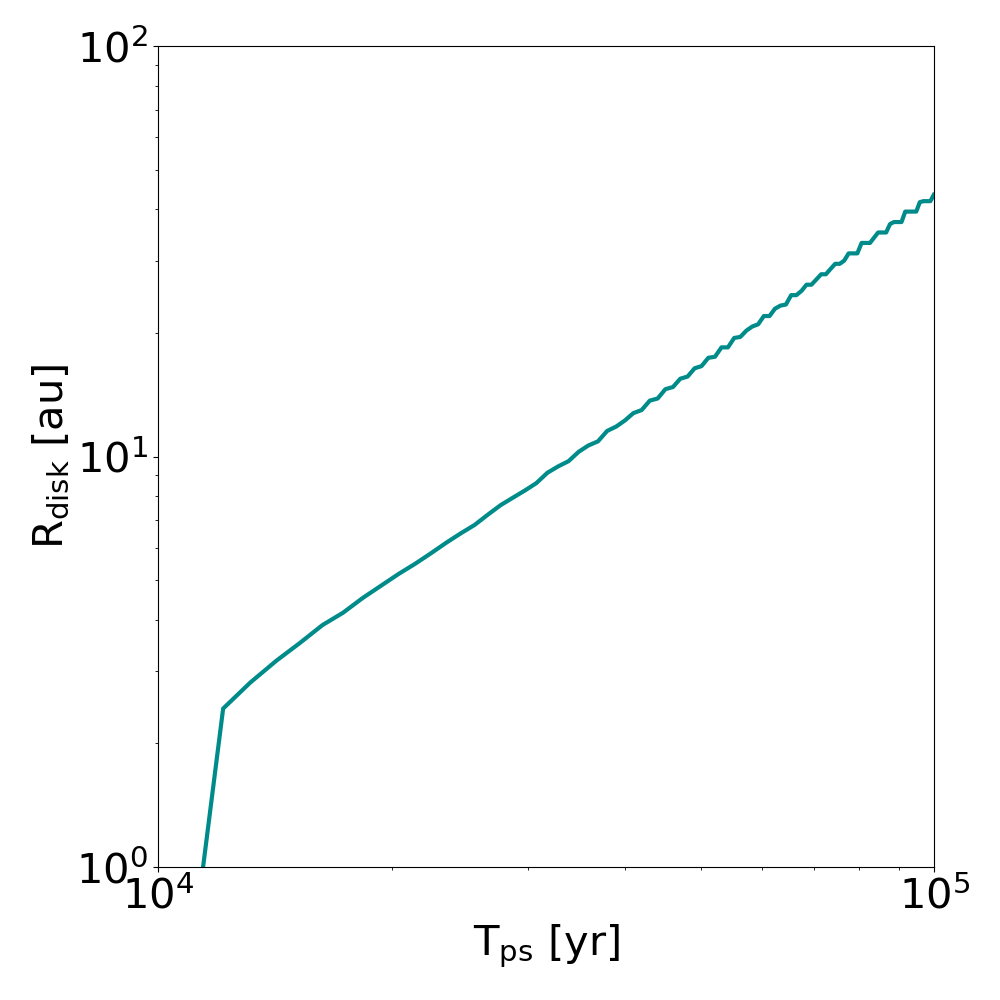}
    \caption{The disk size as a function of time. The vertical axes show disk size and the horizontal axes show time from protostellar formation $\tps$.}
    \label{fig:t_rdisk}
\end{figure}

Figure \ref{fig:t_rdisk} shows the time evolution of the disk size. After disk formation ($\tps>10^4$ years), the disk size monotonically increases with time and, at $\tps=10^5$ years, the disk size becomes $R_{\rm d}=43$ au.
The result that the disk size is several tens of au during the Class 0 phase is consistent with recent survey observations \citep{2024ApJ...973..138H}.
The disk size (and therefore the disk mass) varies slightly vibrationally due to the low resolution at the outer region.

\subsubsection{Comparison between alphaGI and non-alphaGI}
In this section, we compare the results of alphaGI (simulation with $\alpha$ viscosity; fiducial model) and non-alphaGI (simulation without $\alpha$ viscosity).

\begin{figure}[ht]
    \centering
    \begin{minipage}[ht]{\linewidth}
    \includegraphics[width=\linewidth]{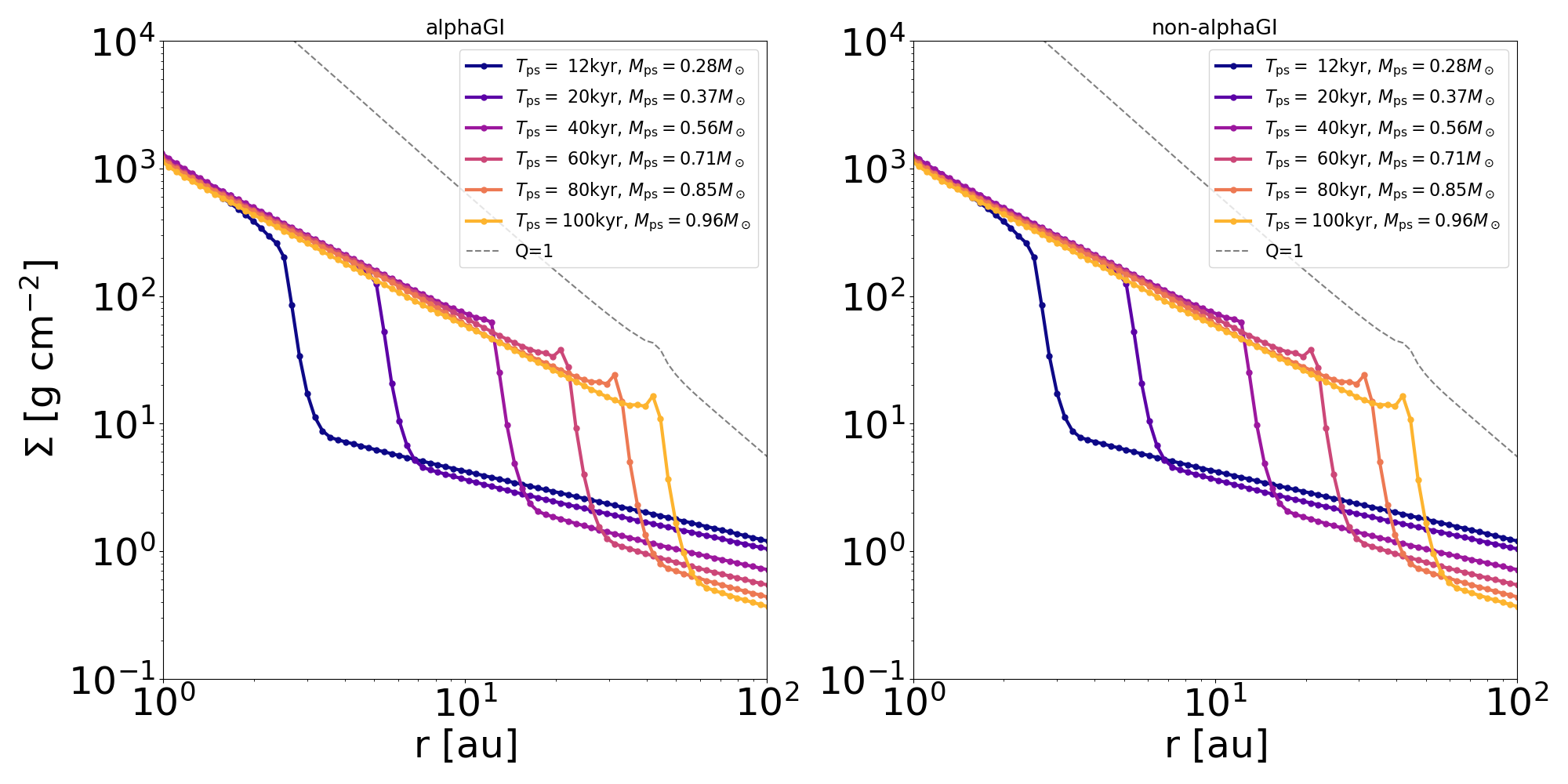}
    \includegraphics[width=\linewidth]{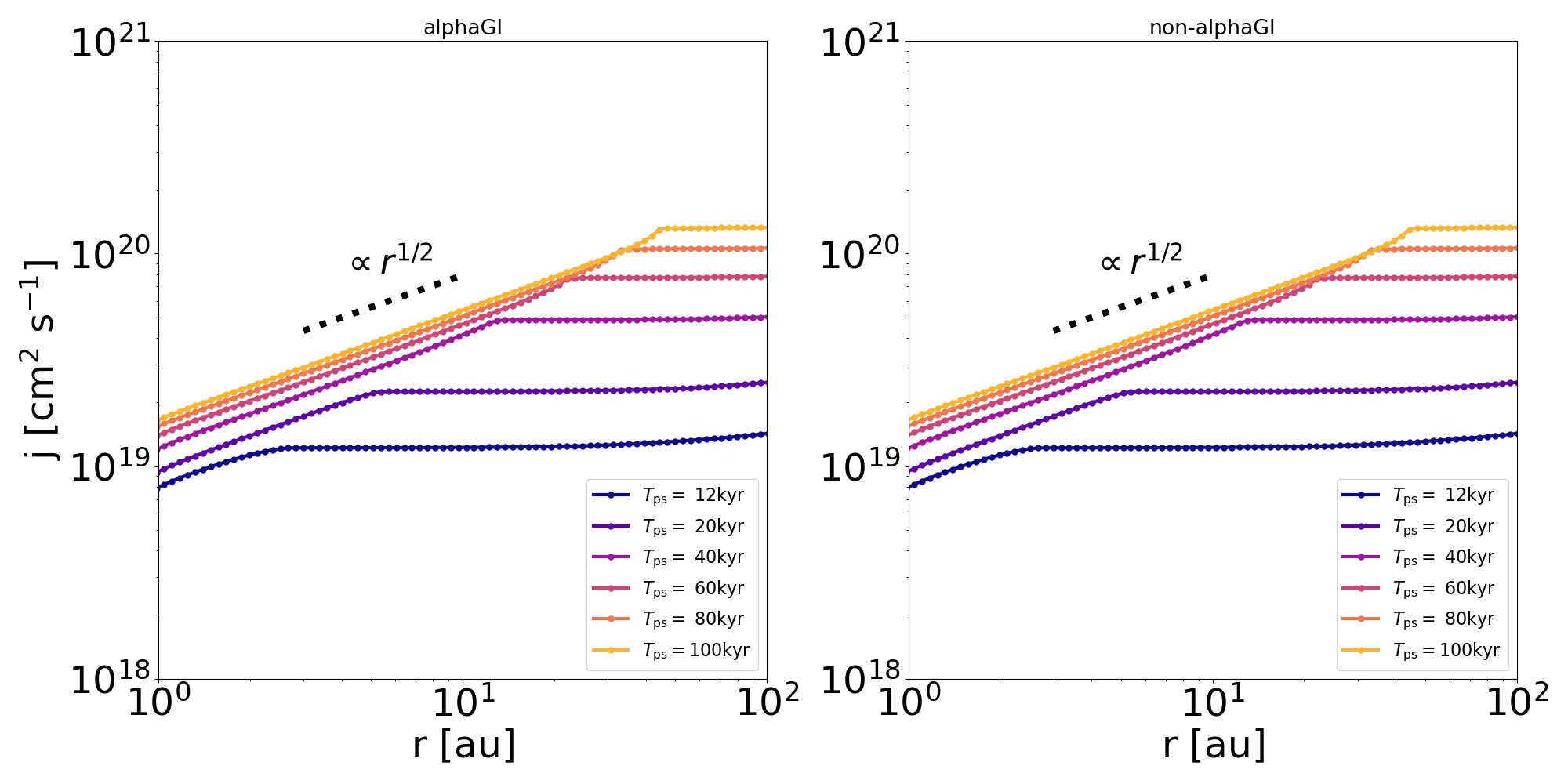}
    \caption{
    The radial profile of the surface density (top row) and specific angular momentum (bottom row). The left column shows the plot of alphaGI, and the right column shows the plot of non-alphaGI. Color is the same as figure \ref{fig:sigma_disk}.The gray dashed line in the surface density profile shows the "$Q=1$" disk. 
    %%% The radial profile of the density (top row) and specific angular momentum (bottom row). The left column shows the plot of alphaGI, and the right column shows the plot of non-alphaGI. Color is the same as figure \ref{fig:sigma_disk}. The black dotted line in the density profile shows the power law that follow the stationary solution according to \cite{2024PASJ...76..674T}.
    The black dotted line in the specific angular momentum profile shows the power law of Keplerian rotation.
    }
    \label{fig:disk_comp}
    \end{minipage}
\end{figure}

Figure \ref{fig:disk_comp} shows the radial profile of the surface density (top row) and the specific angular momentum (bottom row) for alphaGI (left) and non-alphaGI (right).The gray dashed line in the surface density profile shows the surface density of $Q=1$ disks, $\Sigma_Q=\cs\Omega/\pi G$.
%%% Figure \ref{fig:disk_comp} shows the radial profile of the density (top row) and specific angular momentum (bottom row) for alphaGI (left) and non-alphaGI (right). The black dotted line in the density is the power law of the stationary solution according to \cite{2024PASJ...76..674T}, which shows the disk structure growing by magnetic braking and ambipolar diffusion.
The black dotted line in the specific angular momentum shows the Keplerian rotation.

There is no significant difference in the disks between alphaGI and non-alphaGI.
In both models, the disk is formed at $\tps=1.2\times10^4$ years, and its size increases with time.
At $\tps=10^5$ years, the size of these disks is $R_{\rm d}\sim 40$ au.
In each disk, $\Sigma$ is smaller than $\Sigma_Q$ (gray dashed line), indicating that the formed disk is gravitationally stable in both models.
%%% The density follows the power law of the stationary solution ($\rho\propto r^{-21/11}$). This means that they evolve by magnetic braking and ambipolar diffusion. 
The specific angular momentum follows the Keplerian rotation ($j\propto r^{1/2}$) within the disk $r<R_{\rm d}$. 

\begin{figure}[ht]
    \centering
    \begin{minipage}[ht]{\linewidth}
    \includegraphics[width=0.5\linewidth]{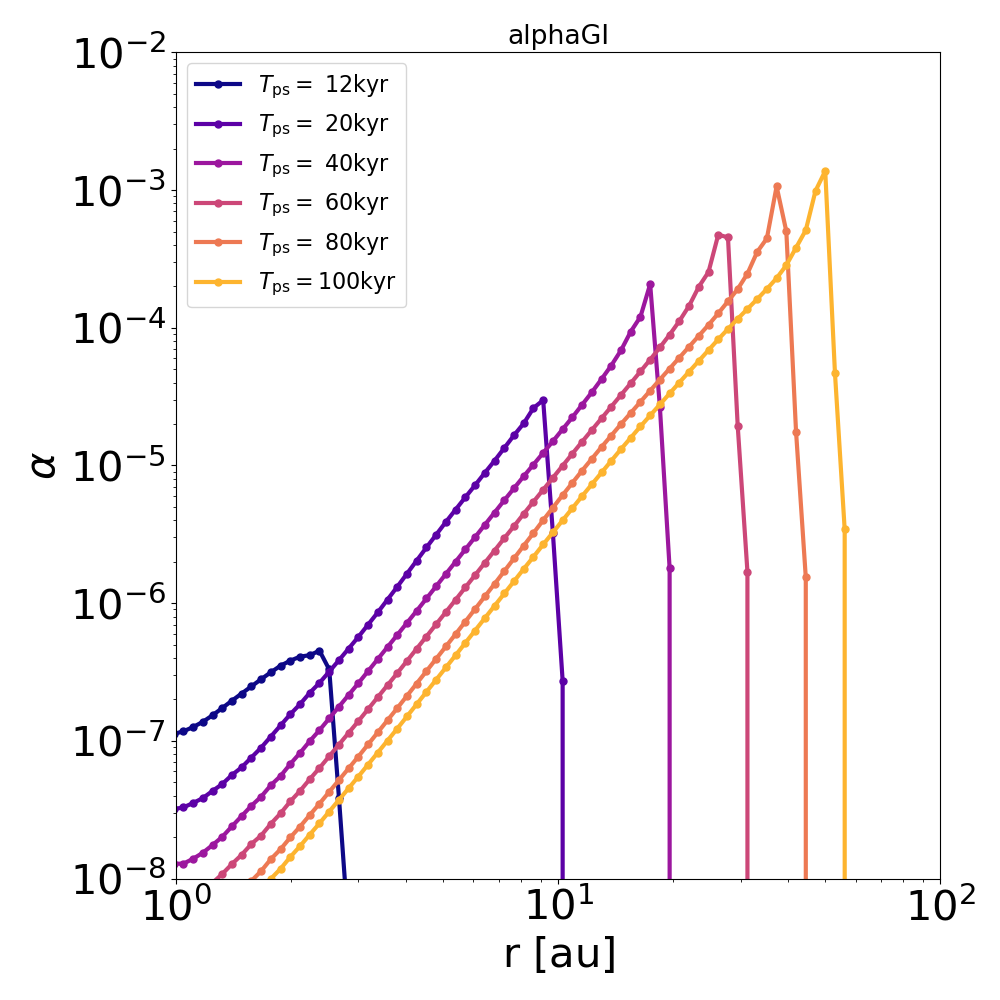}\\
    \includegraphics[width=\linewidth]{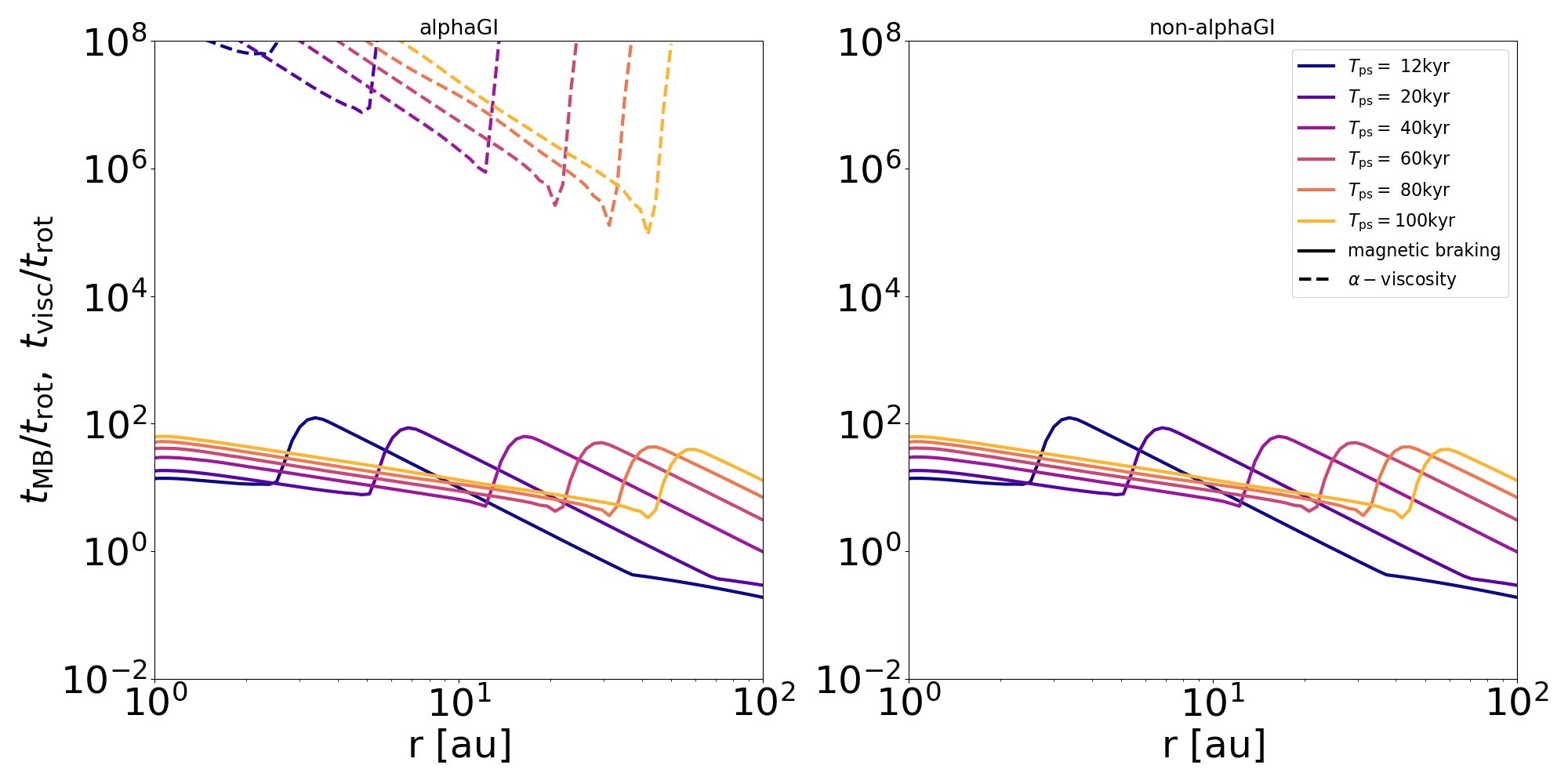}
    \caption{The top panel shows the radial profile of $\alpha$ value in the alphaGI. 
    The bottom panels show the radial profile of angular momentum transfer time scale normalized by rotational timescale $t_{\rm rot}=1/\Omega$.
    The left panel shows the plot of alphaGI and the right panel shows the plot of non-alphaGI. Color is the same as figure \ref{fig:sigma_disk}. Solid and dashed lines show magnetic braking and $\alpha$-viscosity, respectively.}
    \label{fig:infall_rate_disk}
    \end{minipage}
\end{figure}

%Figure \ref{fig:infall_rate_disk} shows the radial profile of angular momentum transfer rate normalized angular velocity $\Omega$.
Figure \ref{fig:infall_rate_disk} shows the $\alpha$ value in alphaGI (top panel) and the angular momentum transfer timescale normalized by the rotational period $t_{\rm rot}=\Omega^{-1}$ (bottom panels).
In the bottom panels, the left panel shows the plot of alphaGI and the right panel shows the plot of non-alphaGI.
Solid and dashed lines show the time scales of magnetic braking $t_{\rm MB}$ and $\alpha$ viscosity (gravitational instability) $t_{\rm visc}$, which are defined as,
\begin{eqnarray}
    % {t_{\rm MB}}^{-1} &=& \frac{1}{2\pi r}\frac{\bphs\bzeq}{\Sigma\Omega},\\
    % {t_{\rm visc}}^{-1} &=& \frac{\alpha \cs H}{r^2}.
    t_{\rm MB} &=& \frac{2\pi r \Sigma \Omega}{\bphs \bzeq}, \\
    t_{\rm visc} &=& \frac{r^2}{\alpha \cs H}.
\end{eqnarray}

In alphaGI, the $\alpha$ value is only $\lesssim 10^{-3}$ even at the outer edge of the disk and decreases sharply toward the inner region (see the top panel of Figure \ref{fig:infall_rate_disk}). Even in the outer edge region where $\alpha$ is largest, the viscous timescale is four orders of magnitude larger than the magnetic braking timescale. 
Therefore, we conclude that the angular momentum transfer due to viscosity is less important than that due to magnetic braking.
%$t_{\rm MB}$ is comparable to $t_{\rm rot}$ by a factor of 10 to 100.

\section{Summary and Discussion} \label{sec:discussion}
In this paper, we developed a one-dimensional non-ideal MHD simulation code that extends the methods of previous studies \citep{2012A&A...541A..35D,1994ApJ...432..720B} to handle protostellar evolution. We demonstrated a long-term simulation of the protoplanetary disk evolution.
The simulations using our code were stably executed from the isothermal contraction phase of molecular cloud cores to $10^5$ years after protostellar formation.

Furthermore, we showed that the disk structure in our simulations agrees well with the steady-state solutions that reproduce the 3D simulation results of previous studies \citep{2024PASJ...76..674T}. This shows that our method, despite being a 1D simulation, possesses predictive power comparable to more realistic 3D simulations.

Furthermore, because it is a 1D simulation, we think that it can easily consider complex physical processes such as dust growth and planetesimal formation, radiation feedback from the central protostar, magnetic disk winds, and magnetic rotation instability (MRI) turbulence at low computational cost.

A particularly important point is that our simulation code solves the equation of vertical magnetic field evolution and it allows us to predict $\alpha$ parameters or similar parameters used in the magnetic disk wind model \citep{2016A&A...596A..74S} when considering them magnetic disk winds and MRI turbulence. This point is a strength of our method compared to previous studies of 1D disk evolution and associated planet formation.

In this study, we focused on investigating the feasibility of the method, so there are several issues that need to be addressed in future studies. In particular, the consideration of magnetic resistivity with charged dust grains, the gas temperature and EOS after protostellar formation, and the introduction of $\alpha$ parameter due to MRI that depends on the vertical magnetic field and magnetic resistivity are urgent issues that need to be investigated early on. We plan to address these issues in our next paper and investigate disk evolution under various initial conditions.

%% Please use the acknowledgment and contribution environments. This will 
%% be anonomyized when the "anonymous" style option is used. 
\begin{acknowledgments}
This work is supported by JSPS KAKENHI grant numbers 24KJ1835, %\textbf
{JST KAKENHI grant numbers JP24K22906 (D.T.),} and JST FOREST grant number JPMJFR2234.
%\textbf
{Simulation code and data will be provided upon request from readers.}
\end{acknowledgments}


\appendix

\subsection{Artificial viscosity}
In this section, we describe the artificial viscosity.
Adopting the model of \citet{1992ApJS...80..753S}, the artificial viscosity term was introduced into the equation of motion \eqref{EoM} for the stability of the simulation.
The artificial viscosity stress $f_{\rm q}$ is given as,
\begin{eqnarray}
    f_{\rm q}
    &=& \frac{\partial q}{\partial r},\\
    q&=&
    \begin{cases}
        C_{\rm vis} \Sigma (v_{{\rm r},i+1}-v_{{\rm r},i})^2 & \text{if }(v_{{\rm r},i+1}-v_{{\rm r},i}<0), \\
        0 & \text{otherwise},
    \end{cases}
\end{eqnarray}
where $i=1,2,...,N-1$, $q$ is the scalar pseudo-viscous pressure, $C_{\rm vis}=3$ is the dimensionless constant which represents the strength of artificial viscosity, and $v_{{\rm r},i}$ and $v_{{\rm r},i+1}$ are the radial velocity of $i$-th and $i+1$-th cell center, respectively. At the innermost($i=0$) and outermost($i=N$) cells, $q=0$.

\subsection{Calculation of Advected quantity}
In this section, we describe the calculation of advected quantities at the cell interface for physical quantities such as surface density $\Sigma$, vertical magnetic field $\bzeq$, momentum $\Sigma\vrad$ and angular mometum $L$.
Those advected quantities are calculated according to \citet{1991ApJ...371..296M} and \citet{1994ApJ...432..720B}.

The advection velocity $v_{\rm adv}$ is the radial velocity at the cell interface.
The $i$-th advection velocity $v_{{\rm adv},i}$ is calculated using the linear interpolation between $i-1$-th and $i$-th cell,
\begin{eqnarray}
    v_{{\rm adv},i}=
    \frac{(r_{i}-\Tilde{r}_i)v_{{\rm r},i-1}+(\Tilde{r}_i-r_{i-1})v_{{\rm r},i}}{{r}_i-{r}_{i-1}},
\end{eqnarray}
where $i=1,2,...,N-1$, $\Tilde{r}_i$ and $r_i$ are the $i$-th position of the cell interface and the cell center, respectively, and $v_{{\rm r},i}$ is the radial velocity at the $i$-th cell center. The boundary conditions at the innermost cell ($i=0$) and the outermost cell ($i=\rm{N}$) are $v_{{\rm adv},i=0}=0,\ v_{{\rm adv},i=N}=0$, respectively. These boundary conditions imply that there is no inflow or outflow of medium into the calculation domain.

%We adopted the upwind differentiation.
%The advection value of $f$, which is a quantity per unit area ($\Sigma$, $\bzeq$, and  $\Sigma\vrad$), is, \citep{1977JCoPh..23..276V}
The advected quantity $f_{{\rm adv}}$ for the quantity $f$ at the cell interface between $i-1$-th and $i$-th cell is given as, 
\begin{eqnarray}
    f_{{\rm adv},i}&=&
    \begin{cases}
        f_{i-1}+(\Tilde{r}_{i}-r_{i-1})\left(\frac{\partial f}{\partial r}\right)_{{\rm adv}, i-1}
        & (v_{{\rm adv},i}\ge0),\\
        f_{i}+(r_{i}-\Tilde{r}_{i})\left(\frac{\partial f}{\partial r}\right)_{{\rm adv},i} & (v_{{\rm adv},i}<0),
    \end{cases}
    \label{eq:fadv}
\end{eqnarray}
where $i=0,1,2,...,N$ and $(\partial f/\partial r)_{{\rm adv},i}$ is given as,
\begin{eqnarray}
    \left(\frac{\partial f}{\partial r}\right)_{{\rm adv},i}&=&
    \begin{cases}
        \frac{(f_{i+1}-f_{i})(f_{i}-f_{i-1})}{(\Tilde{r}_{i+1}-r_{i})(f_{ i}-f_{i-1})+(r_{i}-\Tilde{r}_{i})(f_{i+1}-f_{i})} & (f_{i+1}-f_{i})(f_{i}-f_{i-1})>0,\\
        0 & \text{otherwise}.
    \end{cases}
    \label{eq:dfdz}
\end{eqnarray}

Flux of mass $\mathcal{F}_{m,i}$, magnetic field $\mathcal{F}_{\Phi,i}$, and momentum $\mathcal{F}_{p,i}$  at the $i$-th cell interface are given as,
\begin{eqnarray}
    \mathcal{F}_{m,i}
    &=& \Tilde{r}_{i} \Sigma_{{\rm adv},i} v_{{\rm adv},i}, \label{eq:Fm}\\
    \mathcal{F}_{\Phi,i}
    &=& \Tilde{r}_{i} (\bzeq)_{{\rm adv},i} v_{{\rm adv},i}, \label{eq:Fb}\\
    \mathcal{F}_{p,i}
    &=& \Tilde{r}_{i} (\Sigma \vrad)_{{\rm adv},i} v_{{\rm adv},i}, \label{eq:Fp}
\end{eqnarray}
where $i=0,1,2,...,N$, $\Sigma_{{\rm adv},i}$, $(\bzeq)_{{\rm adv},i}$, and $(\Sigma\vrad)_{{\rm adv},i}$ are the advected quatities at the $i$-th cell interface of the surface density, vertical magnetic field, and momentum calculated from Equation (\ref{eq:fadv}).

The angular momentum is transferred with the mass.
The flux of the angular momentum $\mathcal{F}_{L,i}$ at the $i$-th cell interface is calculated according to \citet{1994ApJ...432..720B} and \citet{2012A&A...541A..35D}.%\citet{1980ApJ...239..968N}.
$\mathcal{F}_{L,i}$ is given as,
\begin{eqnarray}
    \mathcal{F}_{L,i} = j_{{\rm adv},i}^*\mathcal{F}_{m,i},
\end{eqnarray}
where $i=0,1,2,...,N$, $j_{{\rm adv},i}^*$ is the best estimate of the specific angular momentum at the $i$-th cell interface.
The best estimate of the specific angular momentum $j_{{\rm adv},i}^*$ is given as,
\begin{eqnarray}
    j_{{\rm adv},i}^*=
    \begin{cases}
        \Omega_{{\rm adv},i}{\Tilde{r}_i}^2 & \text{if $|\frac{\Omega_{i+1}-\Omega_i}{\Omega_i}|$ smallest},\\
        (\vphi)_{{\rm adv},i}\Tilde{r}_i & \text{if $|\frac{v_{\phi,i+1}-v_{\phi,i}}{v_{\phi,i}}|$ smallest},\\
        j_{{\rm adv},i} & \text{if $|\frac{j_{i+1}-j_i}{j_i}|$ smallest},
    \end{cases} \label{eq:Fj}
\end{eqnarray}
where $i=0,1,2,...,N$, $\Omega_{{\rm adv},i}$, $(\vphi)_{{\rm adv},i}$, and $j_{{\rm adv},i}$ are the advected quantities of angular velocity, rotational velocity, and specific angular momentum at the $i$-th cell interface, respectively.
They are calculated using Equations \eqref{eq:fadv} and \eqref{eq:dfdz}, as is the surface density $\Sigma$, etc.

\subsection{Spatial Derivative}
In this section, we describe the first-order spatial derivatives at the cell center.
The first-order spatial derivative of quantity at the cell center $f$ is calculated according to previous studies of Mouschovias group \citep{1991ApJ...371..296M,1994ApJ...421..561M,1994ApJ...432..720B,2012A&A...541A..35D},
\begin{eqnarray}
    \left(\frac{\partial f}{\partial r}\right)_i
    = a_if_{i-1}+b_if_i+c_if_{i+1},
\end{eqnarray}
where $a_i$,$b_i$, and $c_i$ are the $i$-th derivative coefficients. Here, $f_{i-1}$, $f_i$ and $f_{i+1}$ are the quantities $f$ at the $i-1$-th, $i$-th and $i+1$-th cell center, respectively. Those values are calculated,
\begin{eqnarray}
    a_i &=& -\frac{1}{d_i}\left(
    S_{i+1}+S_i-2{r}_i({r}_{i+1}-{r}_i)
    \right),\\
    c_i &=& -\frac{1}{d_i}\left(
    S_i+S_{i-1}-2{r}_i({r}_i-{r}_{i-1})
    \right),\\
    b_i &=& -(a_i+c_i),\\
    d_i &=& ({r}_i-{r}_{i-1})(S_i+S_{i+1})
    - ({r}_{i+1}-{r}_i)(S_i+S_{i-1}),
\end{eqnarray}
where $S_i\equiv({\Tilde{r}_{i+1}}^2-{\Tilde{r}_i}^2)/2$ is the area of $i$-th cell.

\subsection{Calculation of the Diffusion term}
In our simulation code, the diffusion term such as angular momentum diffusion and magnetic diffusion is calculated using the difference of the first-order spatial derivative at the cell interface. The first-order derivative at the $i$-th cell interface is calculated using linear interpolation,
\begin{eqnarray}
    \left(\frac{\partial f}{\partial r}\right)_{i} = \frac{f_i-f_{i-1}}{{r}_i-{r}_{i-1}}, \label{eq:drvf}
\end{eqnarray}
where $i=1,2,...,N-1$ and $f_i$ is the quantity at the $i$-th cell center.
The boundary conditions at $i=0$ and $i=N$ cell are,
\begin{eqnarray}
    \left(\frac{\partial f}{\partial r}\right)_{i=0} = 0, \quad \left(\frac{\partial f}{\partial r}\right)_{i=N} = \left(\frac{\partial f}{\partial r}\right)_{i=N-1}.\label{eq:drvf0}
\end{eqnarray}
Using Equations \eqref{eq:drvf} and \eqref{eq:drvf0}, the $i$-th diffusion term is calculated as,
\begin{eqnarray}
    \left(\frac{\partial}{r\partial r}\frac{\partial f}{\partial r}\right)_i
    &=& \frac{1}{S_i}\left(\left(\frac{\partial f}{\partial r}\right)_{i+1/2}
    -\left(\frac{\partial f}{\partial r}\right)_{i-1/2}\right).
\end{eqnarray}
Here, we assume $r\partial r=S_i$.

\subsection{Machine Unit}
We normalized the MHD equations by following values,
\begin{eqnarray}
    \Sigma_0
    &=& 2.00\times10^{-2}
    \quad [\rm{g\ cm^{-2}}],\\ % \quad \text{unit of surface density},\\
    c_{s, {\rm 10K}}
    &=& 1.88\times10^4
    \quad [\rm{cm\ s^{-1}}],\\ % \quad \text{unit of velocity},\\
    2\pi G \Sigma_0
    &=& 8.39\times10^{-9} %8.38680\times10^{-9}
    \quad [\rm{cm\ s^{-2}}],\\ % \quad \text{unit of acceleration},\\
    2\pi\sqrt{G}\Sigma_0
    &=& 32.5 %32.4641
    \quad [\rm{\mu G}],\\ % \quad \text{unit of magnetic field},\\
    \left[ t \right]
    &=& \frac{c_s}{2\pi G \Sigma_0} = 2.24\times10^{12} %2.24162\times10^{12}
    \quad [\rm{s}],\\ % \quad \text{unit of time},\\
    \left[ L \right]
    &=& \frac{\cs^2}{2\pi G \Sigma_0} = 4.21\times10^
{16} %4.21424\times10^{16}
    \quad [\rm{cm}],\\ % \quad \text{unit of length},\\
    \left[ m \right]
    &=& \frac{\cs^4}{(2\pi G)^2\Sigma_0} = 2.23\times10^{32} %2.23177\times10^{32}
    \quad [\rm{g}],% \\ % \quad \text{unit of mass},\\
    % \left[ \Omega \right]
    % &=&
    % \frac{2\pi G \Sigma_0}{\cs} = 4.46\times10^{-13} %4.46106\times10^{-13}
    % \quad [\rm{s}^{-1}],% \quad \text{unit of angular velocity},
\end{eqnarray}
where $c_{s, {\rm 10K}}$ is the isothermal sound velocity at 10 K, $G=6.67\times10^{-8}\ \mathrm{g^{-1}\ cm^3\ s^{-2}}$ is the gravitational coefficient in cgs unit.

\bibliographystyle{aasjournal}
\bibliography{export-bibtex}

\end{document}